\documentclass[amsfonts,a4paper,11pt]{article}
\pdfoutput=1

\usepackage{amsmath}
\usepackage{graphicx}
\usepackage{amsfonts}
\usepackage{amssymb} 
\usepackage{bm}
\usepackage{xcolor}
\usepackage{color}
\usepackage{mathtools} 
\usepackage{pifont}
\usepackage{soul}
\usepackage{wrapfig}
\usepackage{caption}
\usepackage{mathrsfs}

\usepackage[colorlinks=true, urlcolor=violet, linkcolor=blue,
  citecolor=red, hyperindex=true, linktocpage=true]{hyperref}

\newcommand{\bc}{\begin{center}}
\newcommand{\ec}{\end{center}}
\newcommand{\bfgr}{\begin{figure}}
\newcommand{\efgr}{\end{figure}}
\def\ba#1{\begin{array}{#1}\displaystyle}
\newcommand{\ea}{\end{array}}

\newcommand{\beq}{\begin{equation}}
\newcommand{\eeq}{\end{equation}}
\newcommand{\beqa}{\begin{eqnarray}}
\newcommand{\eeqa}{\end{eqnarray}}

\newcommand{\n}{\nonumber\\}
\newcommand{\bi}{\begin{itemize}}
\newcommand{\ei}{\end{itemize}}

\def\lt#1{\left#1}
\def\rt#1{\right#1}

\def\b#1{\bar{#1}}
\def\frc#1#2{\frac{#1}{#2}}

\newcommand{\p}{\partial}

\newcommand{\bra}{\langle}
\newcommand{\ket}{\rangle}
\newcommand{\Z}{{\mathbb{Z}}}

\newcommand{\C}{{\mathbb{C}}}

\newcommand{\Or}{{\cal O}}

\newcommand{\si}{\sigma}
\newcommand{\ep}{\epsilon}
\newcommand{\varep}{\varepsilon}

\def\cl#1{\overline{#1}}

\DeclarePairedDelimiter{\floor}{\lfloor}{\rfloor} 

\newcommand{\ig}{\includegraphics}
\newcommand{\lw}{\linewidth}

\begin{document}

\begin{titlepage}

\begin{center}
{\Large {\bf A Monte Carlo method for critical systems in infinite volume: the planar Ising model}

\vspace{1cm}

Victor Herdeiro and Benjamin Doyon}

Department of Mathematics, King's College London\\
Strand, London, U.K.\\
email: benjamin.doyon@kcl.ac.uk

\end{center}

\vspace{1cm}

In this paper we propose a Monte Carlo method for generating finite-domain marginals of critical distributions of statistical models in infinite volume. The algorithm corrects the problem of the long-range effects of boundaries associated to generating critical distributions on finite lattices. It uses the advantage of scale invariance combined with ideas of the renormalization group in order to construct a type of ``holographic'' boundary condition that encodes the presence of an infinite volume beyond it. We check the quality of the distribution obtained in the case of the planar Ising model by comparing various observables with their infinite-plane prediction. We accurately reproduce planar two-, three- and four-point functions of spin and energy operators. We also define a lattice stress-energy tensor, and numerically obtain the associated conformal Ward identities and the Ising central charge.

\end{titlepage}

\tableofcontents

\newpage

\section{Introduction}

Critical models of statistical physics, such as the Ising model at its Curie temperature, offer a unique opportunity to study the emergence of large-scale behaviours. In this context, criticality occurs only on very small subsets of the parameter space of a given model. At such critical points, a second order phase transition takes place and the behaviour of the model changes drastically. Power laws appear in long-range correlations and in divergences of the free energy and susceptibilities, and with the universal amplitudes, they are described by the powerful theoretical framework of conformal field theory (CFT) \cite{ginsparg,df_cft}. These effects are characteristic of the emerging universal collective behaviours, such as the large fluctuating loops separating ordered phases (a precise representation of critical bubbles of the nucleation theory), which have received attention recently especially in the context of conformal loop ensembles \cite{sheffield,werner} and its relation to CFT \cite{doyon_cle}.

Monte Carlo simulations of statistical models are essential in order to explore with precision and from first principles these critical behaviours, and especially the emerging large-scale objects in infinite volume. However, criticality, because of scale invariance, presents its unique numerical challenges. First, critical slow-down increases, in the most basic algorithms, the time necessary to numerically build the full long-range correlations. Second, at criticality, generic boundaries have effects that propagate well into the bulk in the form of conformal boundary conditions \cite{CardyBCFT}. This effectively precludes the direct Monte Carlo numerical simulation of partial, finite-domain configurations in infinite systems. That is, it is hard to simulate the marginals, in the sense of probability theory, of the fluctuating degrees of freedom lying in finite domains with respect to the infinite-volume distribution. We will refer to such marginals as bulk marginals; they encode all local information (all correlation functions and loop distributions inside a given domain) of the infinite-volume model.

In this paper we propose a solution to the latter problem, illustrating it in the context of the planar Ising model. The Ising model is a milestone in the history of statistical physics \cite{ising_history}. While it was first introduced in one dimension by Lenz, Ising's doctoral work \cite{isingLenz}  showed it could not order at large scales on such geometries, and Peierls arguments \cite{peierls} made it the first model shown to exhibit a phase transition for dimensions greater than two. This breakthrough essentially opened the field of phase transitions. Onsager's seminal contribution \cite{onsager} rigorously proved important aspects of the phase transition, in particular the power-law behaviour of the magnetization as function of the external magnetic field, by giving a general solution in two dimensions. Since then, it has been the subject of very extensive research, a great part of the interest being due to its integrability.

In two dimensions and with zero external field, the model is known to go through a second order phase transition for some lattice-dependent critical temperature. At this point, the response functions and correlation length diverge and the system shows an infinite sensitivity to external perturbations. The fluctuations happen on every length scale, this being associated to the emergence of scale invariance. In general, within the framework of quantum field theory (QFT), emeging scale invariance implies invariance under conformal transformations, and combined with locality, leads to an infinite number of constraints on the correlation functions, algebraically encoded via the Virasoro algebra and its representations (the primary operators). This is CFT. These continuum, universal models are characterized, in particular, by their Virasoro central charge $c$, and have been solved (e.g. analytical expressions or precise formulations for correlation functions) under the requirement of minimality \cite{CFTzamolo,ginsparg,df_cft}. The critical Ising lattice is the \textit{simplest} example of such models, with central charge $c=\frac{1}{2}$, and set of primary fields given by $I$ (the identity field), $\si$ (the ``spin'' field -- scaling limit of the binary lattice variable), and $\varep$ (the energy field -- scaling limit of the local energy density).

On the numerical side, the Ising model has also been the subject of extensive research \cite{newman,stAubin1,stAubin2}. It was discovered that near the critical point the autocorrelations of local updates in Monte Carlo Markov chains diverge severely as $\propto L^4$, where $L$ is the linear size of the lattice, whereas the divergence is only $L^2$ far from the critical coupling. This phenomenon is associated to critical slow-down and has been explained as the difficulty for local updates to build the lowest energy excitations of the spectrum. The discovery or engineering of updates based on clusters or lattice flips \cite{swendsenWang,wolff} has been a breakthrough in studying numerically the Ising model at its critical point, and largely solved the problem of critical slow-down.

However, at the critical point, because of the divergence of the correlation length and the response functions, the system also becomes extremely sensitive to its boundary conditions. Essentially, the universal effects of boundary conditions propagate well inside the bulk in the form of the allowed conformal boundary conditions of the CFT model \cite{CardyBCFT}. A direct consequence is that it is presently impossible to sample the infinite volume Ising model even on a restricted finite subdomain: there is no known procedure to obtain, in an efficient fashion, samples representing bulk marginals.

The aim of this paper is two-fold. First, we propose a Markov chain Monte Carlo (MCMC) sampler (which we refer to as a UV sampler) that produces, to good accuracy, samples representing bulk marginals: finite-domain marginals of critical bulk configurations. It does so by using scale invariance in order to generate, up to corrections which we attempt to characterize, a boundary condition that represents the presence of an infinite space beyond it. This may be seen as a ``holographic'' boundary condition, encoding the information of the infinite space beyond the finite domain. The UV sampler does not require the infinite conformal symmetry emerging in two dimensions, hence is generalizable to higher dimensions. It can also be adapted to study the full universal region near criticality (massive QFT).

The well known transfer matrix methods, not based on Monte Carlo sampling, also allow for very accurate studies of various bulk observables in two-dimensional critical models (see for instance \cite{transferMatrix1, transferMatrix2}). However, these are fundamentally based on the infinite conformal symmetry present in two dimensions, are limited by the exponenrial growth of the time as function of the system's size, and seem to be, at least up to now, more limited in the scope of the observables studied, as they do not actually produce planar bulk marginals.

Second, we present an extensive study of local observables that are natural in the CFT context: numerically evaluating some important bulk two-, three- and four-point functions, constructing the holomorphic stress-energy tensor, and verifying the conformal Ward identities. For completeness, we also present a study of the cluster boundaries, extracting some of their main properties. This in part serves to verify the accuracy of the proposed Markov chain, and in part to provide numerical results that may help in further characterizing the Ising critical point.

The paper is organized as follows. In section \ref{sectmarkov} we describe the theory behind the proposed method and its explicit implementation as a UV sampler in the Ising model, and we provide measures of the accuracy of the bulk marginal obtained. In section \ref{sectnumer}, we report on an extensive numerical study of Ising bulk observables, including multi-point functions and the stress-energy tensor, confirming the agreement with known analytical results. We conclude in section \ref{sectconclu}.

\section{A Markov chain for generating bulk marginals}\label{sectmarkov}

Consider the critical Ising model on a finite lattice. As the volume of the lattice is sent to infinity, marginals on fixed finite domains tend to limit distributions: these are the bulk marginals. However, for any large but finite lattice, generic lattice boundary conditions have effects that propagate well into the bulk, because of long-range power-law correlations. Thus the infinite-volume limit is very slowly approached. By renormalization group arguments, the power-law effects of generic boundary conditions are universal: they can be described by conformal boundary conditions of the associated CFT. Informally, this means that a generic lattice boundary condition is seen, from a large distance $r$, as if it were a conformal boundary conditions $b$ of the CFT model at some other, effective distance $r'$. Two characteristics describe these universal effects: the conformal boundary condition $b$ itself, and the ratio $\Lambda$ determining the effective distance $r'= \Lambda r$ at which the conformal boundary condition appears to be. These are the emerging universal properties of the lattice boundary condition.

Clearly, a clever choice of the lattice boundary condition might correspond to a $\Lambda$ that is very large. With such a choice, bulk marginals are easily produced (as with $\Lambda$ large, it is not necessary to actually take the infinite-volume limit of the lattice model). A natural candidate for such a lattice boundary condition is the marginal on a co-dimension-one closed surface of lattice sites that separates the fixed domain from the rest of a very large lattice. Because of the locality of the Ising measure, such a boundary condition fully encodes the rest of the lattice: it is a ``holographic'' boundary condition.

The proposed MCMC sampler uses conformal invariance and the renormalization group in order to implement the above idea, sequentially adjusting the lattice boundary condition so as to make $\Lambda$ larger and larger. The method is an iterative procedure, where each iteration is composed of two steps, which are meant to effectively zoom in, in the sense of the renormalization group, onto a smaller and smaller region around a central point. The first step is a blow-up procedure, which takes a small central region, of linear size $\lambda^{-1}$ times that of the original region, and blows it up to the linear size of the original region. This is the step that zooms in onto a small region, or, equivalently, that sends the original boundary further away. The distribution of the resulting boundary spins is the holographic boundary condition, encoding its exterior. The second step is a re-thermalization, keeping fixed the new boundary spins. This effectively propagates the new boundary condition towards the inside of the region, recovering the information that was absent in the original small central region due to the finite lattice mesh. The implementation of this method using a Markov chain Monte Carlo (MCMC) sampler will be referred to as a UV sampler.

A UV sampler provides a way of performing Monte Carlo simulations of bulk configurations directly on small lattices, without the costly step of thermalizing on large lattices. The method appears to offer a significant advantage over the na\"ive way of generating thermalized configurations on a large lattice and extracting a bulk configuration by discarding all but a small finite sub-lattice. In this paper we concentrate on the two-dimensional Ising model, but similar ideas can be used in any dimensionality.

In this section we give more details concerning the theory of this procedure, its MCMC implementation, and a basic analysis of the quality of bulk marginals obtained.

\subsection{Renormalization-group steps towards the UV fixed point}

\begin{figure}
\bc\ig[width=.4\lw]{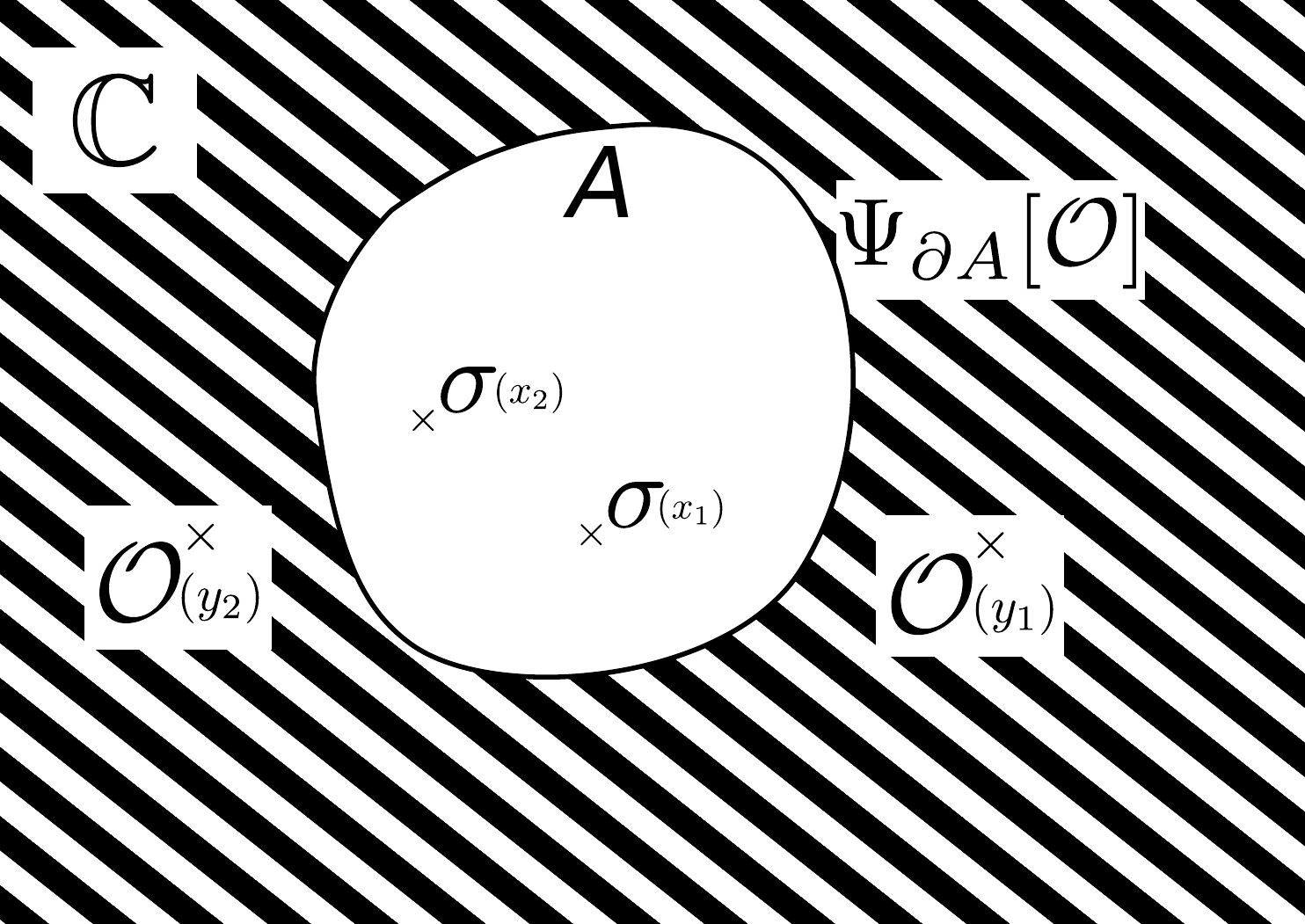}\ec
\caption{Graphical representation of a CFT restricted to a subdomain of the plane. All the outside information (shaded area) is carried by the wave function $\Psi_{\p A}[\Or]$ on the boundary $\p A$.}
\label{figrest}
\end{figure}

In Euclidean CFT on the plane\footnote{Some of the arguments presented here generalize to QFT.}, one may represent, in an appropriate quantization scheme, any region $\C\setminus A$ outside a simply connected region $A\subset \C$, along with any set of local observables $\Or_j(x_j)$ lying outside $A$, by a state $\bra \Psi_{\p A}[\Or]|$ on the boundary $\p A$. This state has enough information to reproduce, for instance, any correlation function with other observables $\sigma_k(y_k)$ lying in $A$:
\beq\label{bdstate}
	\bra \Or_1(x_1)\cdots \Or_n(x_n)\;
	\sigma_1(y_1)\cdots \sigma_m(y_m)\ket
	= \bra \Psi_{\p A}[\Or]| \sigma_1(y_1)\cdots \sigma_m(y_m) |0\ket
\eeq
for $x_j\in \C\setminus \cl A$ and $y_j\in A$ \footnote{To be precise, on the right-hand side of \eqref{bdstate} one uses the representation of the observable $\sigma_j(y_j)$ on the quantization space.}. See Fig. \ref{figrest}.

The state corresponds to restricting on $A$, or ``integrating out'' the exterior $\C\setminus \cl A$. If $A$ is a disk, one may use the radial quantization scheme, where the plane is foliated by concentric circles and time parametrizes their radii. In this scheme, states are seen as linear combinations of field configurations on circles, and $\bra \Psi_{\p A}[\Or]|$ is such a state on the circle $\p A$, obtained from the original correlation function by a decomposition of the identity on $\p A$. But the concept holds for any simply connected region $A$.

Likewise, in a path integral formulation, with measure ${\cal D}\phi \,e^{-S_\C(\phi)}$ over field configurations $\phi$ on the plane, we have:
$$
	\bra \Or_1(x_1)\cdots \Or_n(x_n)\;
		\sigma_1(y_1)\cdots \sigma_m(y_m)\ket
	= \int {\cal D} \phi\; \Or_1(x_1)\cdots \Or_n(x_n)\; \sigma_1(y_1)\cdots \sigma_m(y_m)\, e^{-\beta S_{\C}(\phi)}.
$$
By ultra locality ${\cal D} \phi =  {\cal D} \phi_{\cl A}  \;{\cal D} \phi_{\C\setminus \cl A}$, and by locality $S_{\C}(\phi) =  S_{\cl A}(\phi)+S_{\C \setminus A}(\phi)$ (here $\cl A$ is the closure of $A$), where, because of the microscopic connections (non-ultra locality), both terms in the decomposition of $S_{\C}(\phi)$ depend on the field at the boundary $\p A$. Thus the integral factorizes into:
\beq
\int {\cal D} \phi_{\cl A}\; \sigma_1(y_1)\cdots \sigma_m(y_m)\; e^{-S_{\cl A}(\phi)} \,\Psi_{\p A}[\Or](\phi)
\eeq
where
\beq\label{bdstatepi}
\Psi_{\p A}[\Or](\phi) = 
\int {\cal D} \phi_{\C \setminus \cl A}\; \Or_1(x_1)\cdots \Or_n(x_n)\;e^{-S_{\C \setminus A}(\phi)}
\eeq
carries all the information about operator insertions and the Gibbs measure beyond the boundary, encoding it into a function of the field configuration at the boundary $\phi_{\p A}$. The function $\Psi_{\p A}[\Or](\phi)$ is the ``wave function'' associated to the state $\Psi_{\p A}[\Or]$.

Naturally, in general the state $\Psi_{\p A}[\Or]$ is not a conformally invariant boundary state: it is not invariant under conformal transformations on $A$. With trivial operators $\Or_j={\bf 1}$, it is, however, M\"obius invariant.

The above can also be done on any domain of definition $C$ instead of the plane $\C$, and with any boundary state on $\p C$. The state $\Psi_{\p A}$ will then depend not only on $\Or_j$ but also on the domain $C$ and the boundary conditions on $\p C$. Let us specialize to trivial exterior observables $\Or_j={\bf 1}$. We write $\Psi_{\p A}[\Phi]$ for the state on $\p A$ characterized by the ``exterior information'' $\Phi$ (the information outside $A$): this is the boundary state at $\p C$, which we understand as implicitly containing the information of the domain of definition $C$ itself. On a simply connected domain $C$ with some boundary state $\Phi_{\p C}$, we have $\bra \sigma_1(y_1)\cdots \sigma_m(y_m)\ket_{C} = \bra \Phi_{\p C}| \sigma_1(y_1)\cdots \sigma_m(y_m)|0\ket$, and then the restriction to $A$ can be written as
\beq\label{bdstate2}
	\bra \Phi_{\p C}| \sigma_1(y_1)\cdots \sigma_m(y_m)|0\ket
	= \bra \Psi_{\p A}[\Phi_{\p C}]| \sigma_1(y_1)\cdots \sigma_m(y_m) |0\ket.
\eeq
The following ``projection'' property is a simple consequence of these definitions:
\beq\label{restr}
	\Psi_{\p A}[\Psi_{\p C}[\Phi]] = \Psi_{\p A}[\Phi]\quad (A\subset C).
\eeq
The projection property is very natural from the viewpoint of the path integral formulation, where $\Psi_{\p A}[\Phi](\phi)$ can be evaluated by two embedded path integrals on complementary subdomains.

Let us now consider scale transformations, and restrict ourselves to convex domains. The scale transform $\lambda\cdot \Psi_{\p A}[\Phi]$ of the state $\Psi_{\p A}[\Phi]$ by a factor $\lambda$ is a state on $\lambda \p A$, of the form $\Psi_{\lambda \p A}[\Phi']$. The scale transform is defined by requiring scale invariance $\bra \lambda\cdot \Psi_{\p A}[\Phi]|\lambda \cdot \sigma_{1}(x_1)\cdots \lambda \cdot \sigma_{k}(x_k)|0\ket = \bra \Psi_{\p A}[\Phi]|\sigma_{1}(x_1)\cdots \sigma_{k}(x_k)|0\ket$. The above then simply gives
\beq\label{psia}
	\lambda\cdot \Psi_{\p A}[\Phi] = \Psi_{\lambda \p A}[\lambda \cdot \Phi]
\eeq
(where for instance a domain transforms as $\lambda\cdot C = \lambda C$, and scaling fields as $\lambda\cdot\Or_j(x) = \lambda^{d_j}\Or_j(\lambda x)$, with $d_j$ the scaling dimension of $\Or_j$). Rewriting, we have
\beq\label{psic}
	\Psi_{\p C}[\lambda\cdot \Phi] = \lambda \cdot \Psi_{\lambda^{-1}\p C}
	[\Phi].
\eeq

Let us define the operation
\beq\label{R}
	R_\lambda[\cdots] := \lambda\cdot\Psi_{\lambda^{-1}\p C}[\cdots].
\eeq
This operation is a restriction to a smaller domain, $\Psi_{\lambda^{-1}\p C}[\cdots]$, followed by a scaling by a factor $\lambda$. The map $R_\lambda[\cdots]$ acts on, and generates, states on $\p C$. Let us take $\lambda>1$. We now show that its $N^{\rm th}$ power has the effect of scaling out the domain by a factor $\lambda^N$:
\beq\label{RNC}
	R_\lambda^N[C] = \Psi_{\p C}[\lambda^N C].
\eeq
Here, for lightness of notation, the symbol $C$ in the square brackets represents some (unspecified) {\em conformal} boundary condition on the boundary $\p C$ of the domain of definition $C$. Clearly, \eqref{RNC} is true at $N=0$. Assume it holds for some $N\geq 0$. Then
\beqa
	R_\lambda^{N+1}[C]  &=& \lambda \cdot \Psi_{\lambda^{-1} \p C}[\Psi_{\p C}[\lambda^N C]] \n
	\mbox{(by \eqref{psic})}
	&=& \Psi_{\p C}[\lambda\cdot \Psi_{\p C}[\lambda^N C]] \n
	\mbox{(by \eqref{psia})}
	&=& \Psi_{\p C}[\Psi_{\lambda \p C}[\lambda^{N+1} C]] \n
	\mbox{(by \eqref{restr})}
	&=& \Psi_{\p C}[\lambda^{N+1} C],\label{proofR}
\eeqa
which shows \eqref{RNC} by induction on $N$. Thus,
\beq\label{limR}
	\lim_{N\to\infty} R_\lambda^{N}[C] = \Psi_{\p C}[\C],
\eeq
where on the right-hand side, the symbol $\C$ in the square brackets represents the exterior information of the domain of definition being the infinite plane. Therefore we see that the infinite power of $R_\lambda$ on $C$ produces a boundary state on $\p C$ representing the integrating out of the exterior on the plane $\C$.

\begin{figure}\bc
\ig[width=.7\lw]{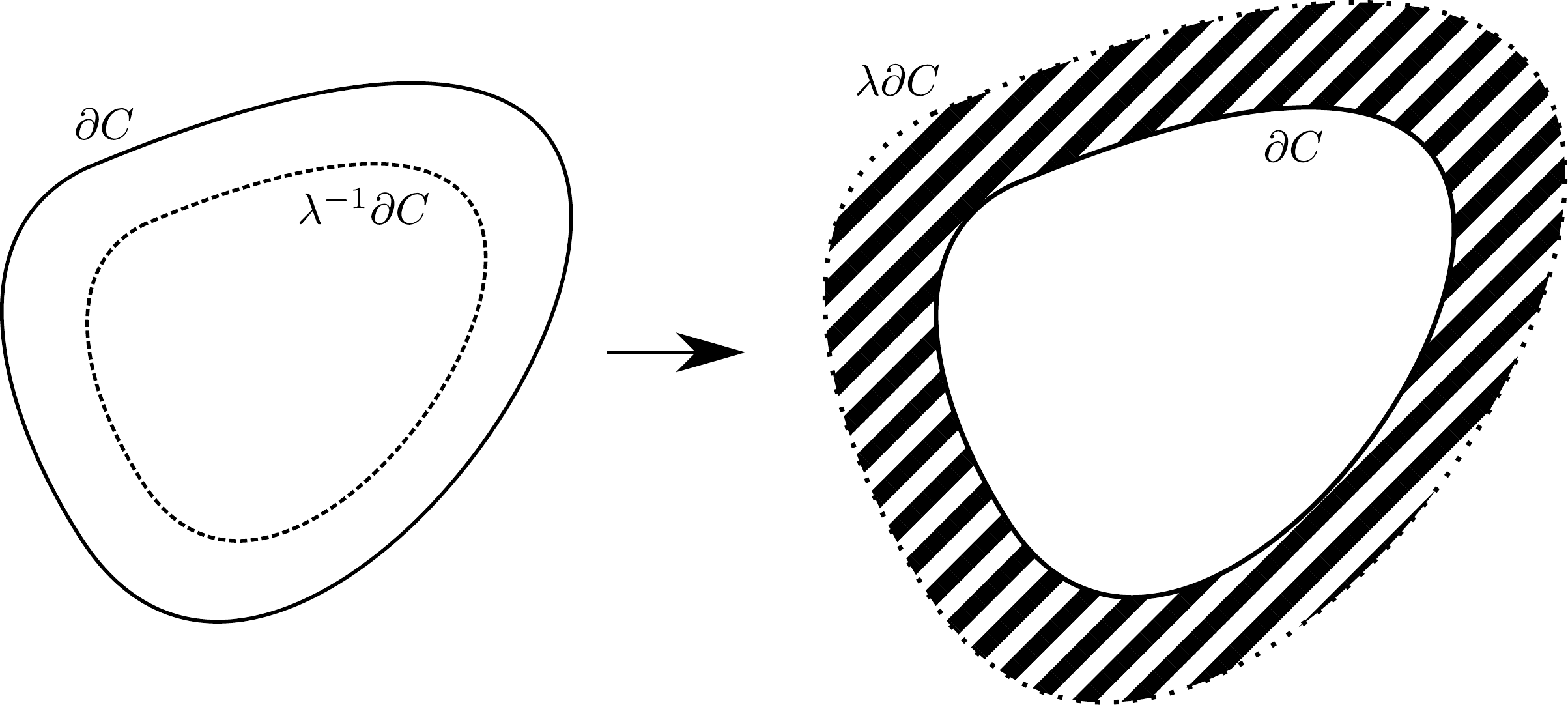}
\caption{Picture of the $R_\lambda$ operation: first $\Psi_{\lambda^{-1} \p C}$ restricts to a small domain, and then this is scaled by $\lambda$. Effectively, the shell $(\lambda \p C,\p C)$ is integrated over. Repeating this step sends the initial domain $C$ to the plane. This can also be viewed as a zoom onto $\lambda^{-1} C$, hence a flow towards the UV.}
\label{figRG}
\ec\end{figure}

The above has a clear interpretation. The map $R_\lambda$ represents, as a mapping of states on $\p C$, the operation of scaling up the domain of definition by a factor $\lambda$, thus sending further away the domain boundary and increasing the space exterior to $C$. It can be seen as a renormalization-group (RG) step on states on $\p C$: $R_\lambda$ tells us how the wave function of a state transforms when integrating out a shell of finite thickness near the boundary of the region, and scaling out the result back to the original domain (see Fig. \ref{figRG}).
This RG step flows towards the ultra-violet fixed point, as the boundary $\lambda^N \p C$, which has the meaning of an infrared cutoff, is sent to infinity while the observables in $C$ are kept invariant; by scale invariance, this is equivalent to zooming onto a smaller and smaller region inside the original domain $C$.

\begin{figure}\bc
\ig[width=.4\lw]{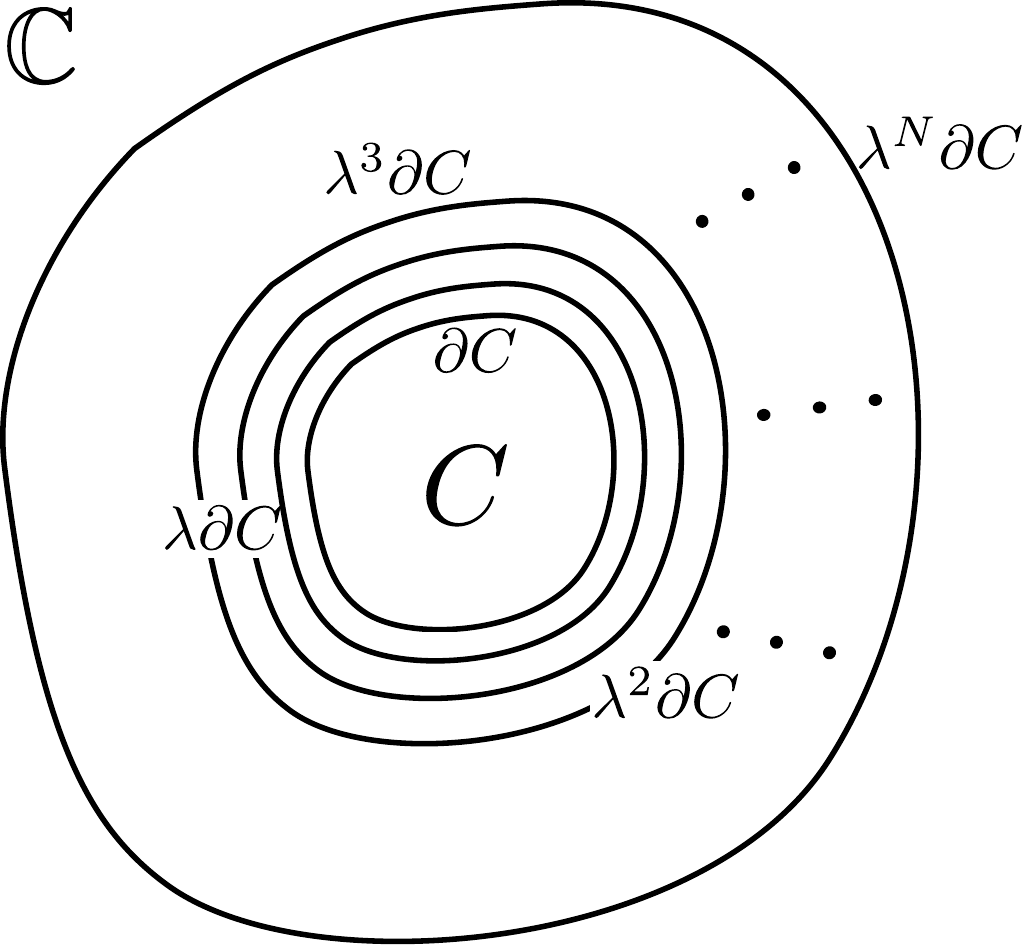}
\caption{Pictorial representaion of the chain of RG steps for a finite number N of dilations. In the limit, $N\rightarrow \infty$, integrating the shells ($\lambda^i \p C,\lambda^{i-1}\p C)$ for $i=N,\dots,1$, is equivalent to integrating out the outside of $C$ up to infinity.}
\label{figchain}
\ec\end{figure}

We therefore have an infinite chain of states on $\p C$, starting with some conformal boundary condition $\Phi^{(0)} = C$, and whose step is an RG transformation towards the ultraviolet (see Fig. \ref{figchain}):
\beq\label{chainCFT}
	\cdots \mapsto \Phi^{(j)} \stackrel{R_\lambda}\mapsto \Phi^{(j+1)}
	\mapsto \cdots
\eeq
This chain converges to the state $\Phi_\infty = \Psi_{\p C}[\C]$. This is the state obtained by integrating out everything up to infinity, thus a bulk marginal. The chain allows us to reach this marginal from a finite domain by performing (an infinite number of) finite-scale RG transformations.

\subsection{Towards the UV fixed point on lattices: blow-up and re-thermalization}

The above CFT description of boundary states representing exterior data has a counterpart in local statistical models. As an example, consider the two-dimensional Ising model. Its configuration space is the set of all functions ${\bf j}\mapsto \si_{\bf j}$ from a regular planar lattice embedded into the plane (say the triangular lattice), to the set $\{1,-1\}$. The (unnormalized) measure on a configuration is
\beq\label{Ising}
	e^{\beta \sum_{({\bf j}, {\bf k})\in{\cal E}} \si_{\bf j} \si_{\bf k}},
\eeq
where ${\cal E}$ is the set of edges of the lattice (and $({\bf j},{\bf k})$ is the edge joining $\bf j$ to $\bf k$). For any domain (open set) $A\subset \C$ (we assume that domain boundaries do not intersect any vertex of the lattice), we denote by $\si_A$ the restriction of $\si$ to the set of all vertices lying in $A$, and by $\si_{\p A}$ the restriction of $\si$ to the set of ``boundary vertices'' $\{{\bf j}\in \C\setminus  A : \exists\;{\bf k}\in A|({\bf j},{\bf k})\in{\cal E}\}$ (a boundary vertex is a vertex outside $A$ which is connected to at least one vertex in $A$). We will also use the notation $\si_{\cl A} = (\si_A,\si_{\p A})$. Because of the regular, planar structure of the lattice, the set of boundary vertices indeed stays uniformly near to the boundary $\p A$ for any domain (open set) $A$. We also denote by ${\cal E}_A$ the set of edges that have nonzero intersection with $A$. We may then define the probability measure on any domain $A$, with boundary conditions fixing $\si_{\p A}$, by $P_A(\si_A|\si_{\p A}) \propto e^{\beta\sum_{({\bf j},{\bf k})\in{\cal E}_A} \si_{\bf j}\si_{\bf k}}$, and the induced probability measures on subsets $B\subset A$ is $P_A(\si_B|\si_{\p A}) = \sum_{\si_{A\setminus \cl B}} P_A(\si_A|\si_{\p A})$.

Let $C\supset A$ be two domains, and consider the restricted probability $P_C(\si_A|\si_{\cl C\setminus \cl A})$: the probability of $\si_A$ given $\si_{\cl C\setminus \cl A} = (\si_{C\setminus \cl A}, \si_{\p C})$. Since the measure {\em factorizes on the set of edges} (that is, the measure is a product of factors, one for each edge), this satisfies a ``domain Markov property": $P_C(\si_A|\si_{\cl C\setminus \cl A}) = P_A(\si_A|\si_{\p A})$. That is, the information of $\si_{\p A}$, rather than the full $\si_{\cl C\setminus \cl A}$, is sufficient in order to deduce the marginal on $A$. Then, if $\Or$ is a random variable supported on $C\setminus \cl A$ and ${\cal A}$ a random variable supported on $A$, the expectation of $\Or{\cal A}$ on the domain $C$ with boundary condition $\si_{\p C}$ can be written as
\beqa
	\mathbb{E}_C(\Or{\cal A}|\si_{\p C}) &=&
	\sum_{\si_C} P_C(\si_C|\si_{\p C}) \Or(\si_{C\setminus \cl A})
	{\cal A}(\si_A)\n
	&=& \sum_{\si_C} P_C(\si_A|\si_{\cl C\setminus \cl A})
	P_C(\si_{C\setminus \cl A}|\si_{\p C})
	\Or(\si_{C\setminus \cl A}){\cal A}(\si_A) \n
	&=& \sum_{\si_{\cl A}}
	{\cal A}(\si_A) 
	P_A(\si_A|\si_{\p A}) \Psi_{\p A}[\Or,C](\si_{\p A})
\eeqa
where
\[
	\Psi_{\p A}[\Or,C](\si_{\p A}) = \sum_{\si_{C\setminus \cl A}\setminus \si_{\p A}}
	\Or(\si_{C\setminus  \cl A}) P_C(\si_{C\setminus \cl A}|\si_{\p C}).
\]
That is, $\mathbb{E}(\Or{\cal A})$ can be evaluated as an expectation on $A$, with boundary condition on $\p A$ determined by the ``wave function'' $\Psi_{\p A}[\Or,C]$. This is the lattice counterpart of \eqref{bdstate}, or of its path-integral version \eqref{bdstatepi}\footnote{Note that in this discrete version of the path integral formulation of the boundary state, the subtlety about $S_{\cl A}$ and $S_{\C\setminus A}$ both depending on the boundary field becomes somewhat clearer.}.

This naturally suggests that we define the expectation with boundary condition $\Phi_{\p C}$ (a function of $\si_{\p C}$) as
\beq
	\mathbb{E}_C({\cal A}||\Phi_{\p C}) = \sum_{\si_{\cl C}} {\cal A} (\si_C) 
	P_C(\si_C|\si_{\p C}) \Phi_{\p C}(\si_{\p C}).
\eeq
We then have the counterpart of \eqref{bdstate2}, for any variable ${\cal A}$ lying in $A$,
\beq
	\mathbb{E}_C({\cal A}||\Phi_{\p C})
	= \mathbb{E}_C({\cal A}||\Psi_{\p A}[\Phi_{\p C}])
\eeq
where the new wave function is
\beq
	\Psi_{\p A}[\Phi_{\p C}](\si_{\p A})
	= \sum_{\si_{\cl C\setminus \cl A}|\si_{\p A}}
	P_{C}(\si_{C\setminus \cl A}|\si_{\p C})
	\Phi_{\p C}(\si_{\p C}).
\eeq
It is clear, from this, that the projection property \eqref{restr} holds.

The operation of dilation in the discrete case is ambiguous. Let $\lambda>1$. Given a measure for the subset of the lattice lying in $A$, a natural definition would be to construct a measure for the subset of the lattice lying in $\lambda A$ by attributing spins $\si'_{\lambda A}$ from $\si_A$ as $\si_{[\lambda {\bf j}]}' = \si_{\bf j}$ for all $\bf j$ on the lattice, where $[\lambda {\bf j}]$ represents the site on the lattice nearest to the coordinate $\lambda {\bf j}$. The main problem\footnote{Another potential problem is that $[\lambda {\bf j}]$ may lie outside $\lambda A$, but this small perturbation of the boundary is not important.} is that this leaves ``holes'': unattributed sites. This is a fundamental property of discrete systems: there is more information on larger region, hence the operation of scaling by a factor greater than one needs to be supplemented by additional information\footnote{The inverse definition $s'_{\bf j} = \si_{[\lambda^{-1}{\bf j}]}$ does not leave holes, but it is effectively just a special prescription for filling in the holes, not expected to be particularly good as it tends to increase correlations (creating small-scale lumps).}, which affects the definition of scaling. Certainly, all correlations of spins on the attributed sites are unambiguously the same, up to scaling, as those of their pre-image, but other correlations depend on the choice of filling prescription. In order not to affect large-distance behaviours, the filling prescription $\si_A\to \si_{\lambda A}'$ should be local ($\si_{\bf j}'$ determined in terms of $\si_{[\lambda^{-1}{\bf j}]}$ and few of its neighbours). Formally, we need such a prescription only for the boundary spins, as we need to describe dilations of states. Given a prescription that determines $\si_{\lambda \p A}'$ in terms of $\si_{\p A}$ (or in terms of the spins on a few-site neighbourhood of $\p A$), we define the scaled state as
\beq\label{latsca1}
	(\lambda \cdot \Psi_{\p A}[\Phi]) (\si_{\lambda \p A}')
	= \Psi_{\p A}[\Phi](\si_{\p A}).
\eeq

Suppose that the measure is chosen to be critical (for instance, the inverse temperature $\beta$ in the Ising model measure \eqref{Ising} is taken at its critical value, which depends on the lattice chosen). In this case, there is indeed scale invariance in the lattice model {\em for large-distance observables}: scale invariance is a property of emerging collective behaviors. Therefore, although it is not expected that the wave function $\Psi_{\p A}[\Phi](\si_{\p A})$, as a function of the spins $\si_{\p A}$, satisfy strict scale invariance properties, the large-distance features of the function are scale invariant. In the context of evaluating observables that do show scale-invariant properties, we expect these large-distance features to dominate. Since the prescription $\si_{\p A}\to \si_{\lambda \p A}'$ only affects local features, lattice scale invariance implies
\beq\label{latsca2}
	\Psi_{\lambda \p A}[\lambda\cdot \Phi](\si_{\lambda \p A}')
	=\Psi_{\p A}[\Phi](\si_{\p A}) + \mbox{small-scale corrections}.
\eeq
In terms of Fourier modes along $\p A$, we would expect the small-scale corrections only to have significant Fourier components at large frequencies.

The combination of \eqref{latsca1} and \eqref{latsca2} gives rise to \eqref{psia}, up to small-scale corrections. Thus, defining the RG operation $R_\lambda$ in the lattice case as in \eqref{R}, the derivation \eqref{proofR} can be reproduced, and we conclude that the chain \eqref{chainCFT} is a Markov chain for generating marginal configurations of a bulk critical lattice, with microscopic modifications due to the small-scale corrections.

The numerical implementation of each iteration of the Markov chain \eqref{chainCFT} involves two steps.
\bi
\item[1.] {\em Re-thermalization.}
Say the state $\Phi_{\p C}$ is reached after $N$ iteration. For the next iteration, first the map $\Phi_{\p C}\mapsto \Psi_{\lambda^{-1}\p C}[\Phi_{\p C}]$ is performed, and this involves a re-thermalization step. Indeed, this requires extracting a spin configuration on $\lambda^{-1} \p C$, and thus requires propagating the boundary condition from $\p C$ to $\lambda^{-1}\p C$. This can be performed by thermalizing the model in $C$ with boundary $\Phi_{\p C}$: generating a typical configuration in $C$ with that boundary condition, from which the spin configuration on $\lambda^{-1}\p C$ can be extracted. The thermalizing process may be done by a multitude of available numerical algorithms, and the details will be explained in subsection \ref{ssectimple}. This part of the numerical implementation is time consuming, but can be as precise as necessary, its accuracy being essentially limited only by the small fluctuations associated to microscopic thermalization.
\item[2.] {\em Blow-up.} Second, a blow-up step is performed. This step is the implementation of $\Psi_{\lambda^{-1}\p C}[\Phi_{\p C}]\to \lambda \cdot \Psi_{\lambda^{-1}\p C}[\Phi_{\p C}]$. This is very fast, but involves an ad-hoc prescription for filling-in the holes, as discussed above. Here we use the extra information gained by thermalizing on all of $C$ in the previous step, and implement this blow-up not only for the boundary spins on $\lambda^{-1}\p C$ but for all spins on $\lambda^{-1}C$. The result is a configuration on $C$ which has the correct boundary condition $\lambda\cdot \Psi_{\lambda^{-1}\p C}[\Phi_{\p C}]$, and which is relatively near to the correct thermalization on the interior of $C$, preserving the correct large-distance correlations. This gives an initial condition in $C$ that accelerates the subsequent re-thermalization step in the next iteration of the chain. The subsequent re-thermalization step is then mostly meant to reconstruct the short distances correlations damaged by the introduction of the holes.
\ei

It is a nontrivial matter to understand and assess the effects of the small-scale corrections and of the prescription for filling-in the holes in the above algorithm. We discuss these aspects in subsection \ref{sec:deltaEexponentsV2}.

\subsection{Implementation}\label{ssectimple}

In this section, we present our proposal for a Markov chain Monte Carlo sampler implementing the function $R_{\lambda} [ \cdot ]$ introduced above in the specific case of the lattice Ising model. For definiteness, we choose the triangular lattice, and the sublattice $C$ is rectangular shaped, of size $(h,v)$. We note that we have also implementated the algorithm and performed verifications on the square lattice.

Our procedure is the following:
\begin{itemize}
\item At the starting point the system has all spins pointing up.
\item We treat the finite lattice as a torus and apply periodic boundary Wolff cluster flips \cite{wolff}. With no external magnetic field the unique coupling is the parameter $\beta$ in \eqref{Ising}, which is taken to be the infinite-volume critical value, here equal to
\beq\label{critcoupl}
	\beta = \beta_c^{\text{triangular Ising}} = \frac{1}{2}\;
    {\rm arcsinh(\frac{1}{\sqrt{3}})} \sim 
    0.274653072\dots
\eeq
Our simulation uses a value close enough to have a correlation length $\xi \gg L$. This step is meant to bring the lattice close to the critical bulk with relatively little computational effort.

\item We then start a cycle by applying a dilation procedure. The ``discrete" dilations will expand a central fraction of the lattice to the whole lattice. For a dilation by a factor $\lambda$, the spin assigned to the site of coordinates ${\bf j} = ({j_1},{j_2})$ will be the value of the previous spin at location:
\beq \label{DiscreteDilFormula}
\floor{\frac{h}{2} + \frc1\lambda\lt({j_1}-\frac{h}{2}\rt) + \delta}\quad , \quad\floor{\frac{v}{2} + \frc1\lambda\lt({j_2}-\frac{v}{2}\rt) + \delta}
\eeq
with $\floor{x}$ the floor function, and where $\delta \in (0,1)$ is a uniformly distributed random parameter fixed for a given dilation (see the motivation in the caption of Fig. \ref{FigLargeLattDil}). Graphically and at the microscopic level, the operation looks like Fig. \ref{FigSmallLattDil}.
\bfgr\bc
\includegraphics[width=.8\lw]{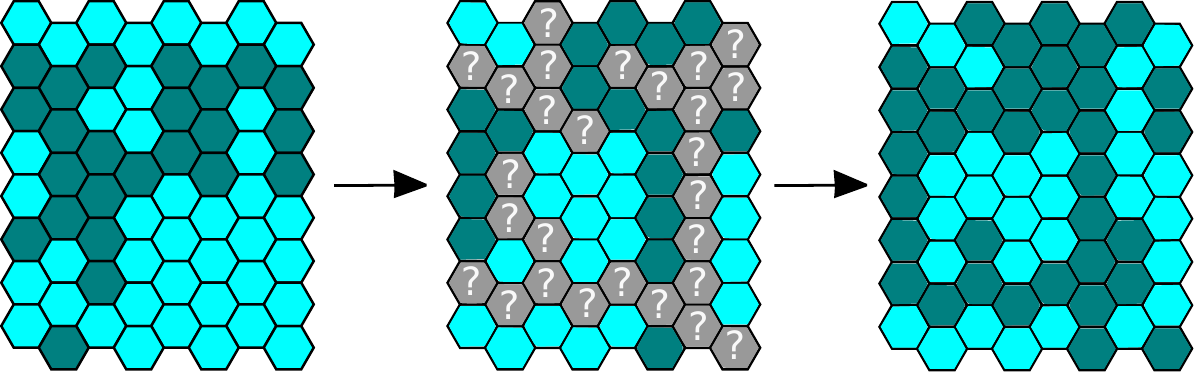}
\caption{\small{From left to right, we apply a dilation of parameter $\lambda = \frac{5}{4}$ between the first and the second graph. The question marks in this middle graph are the holes left behind, and the final graph is the end result after applying the assignation procedure described in the text. The holes have been filled by a heat-bath weighted random assignation, using the information of the neighbour spins of each hole. In the cases where holes are neighbours, the heat bath excludes the edges joining two holes. This is a temporary modification of the lattice Hamiltonian. For holes on the inside, the effect is expected and checked to be harmless: they are dynamical with respect to the lattice flips of the re-thermalization step. For the holes on the border, the heat-bath assignment is in any case approximative, as it misses the forgotten spins beyond the boundary. This induces a departure in the energy density $\bra {\cal H} \ket$ whose profile and dependence on $\lambda$ is studied in \hyperref[sec:deltaEexponents]{subsection 2.6}}}
\label{FigSmallLattDil}
\ec\efgr
On much larger lattices, this discrete dilation is illustrated by Fig. \ref{FigLargeLattDil}.
\bfgr\bc
\ig[width=.32\lw]{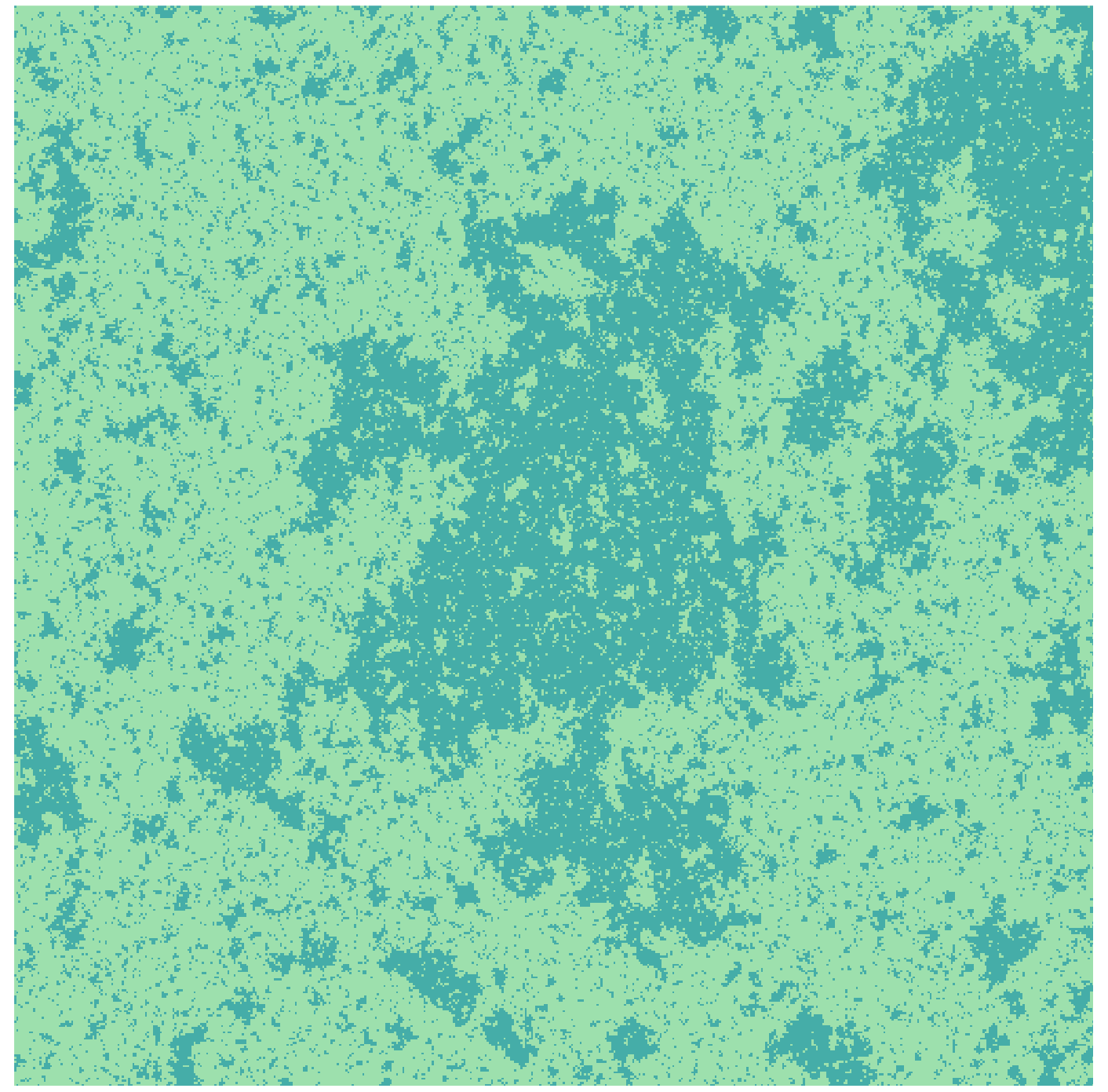}
\ig[width=.32\lw]{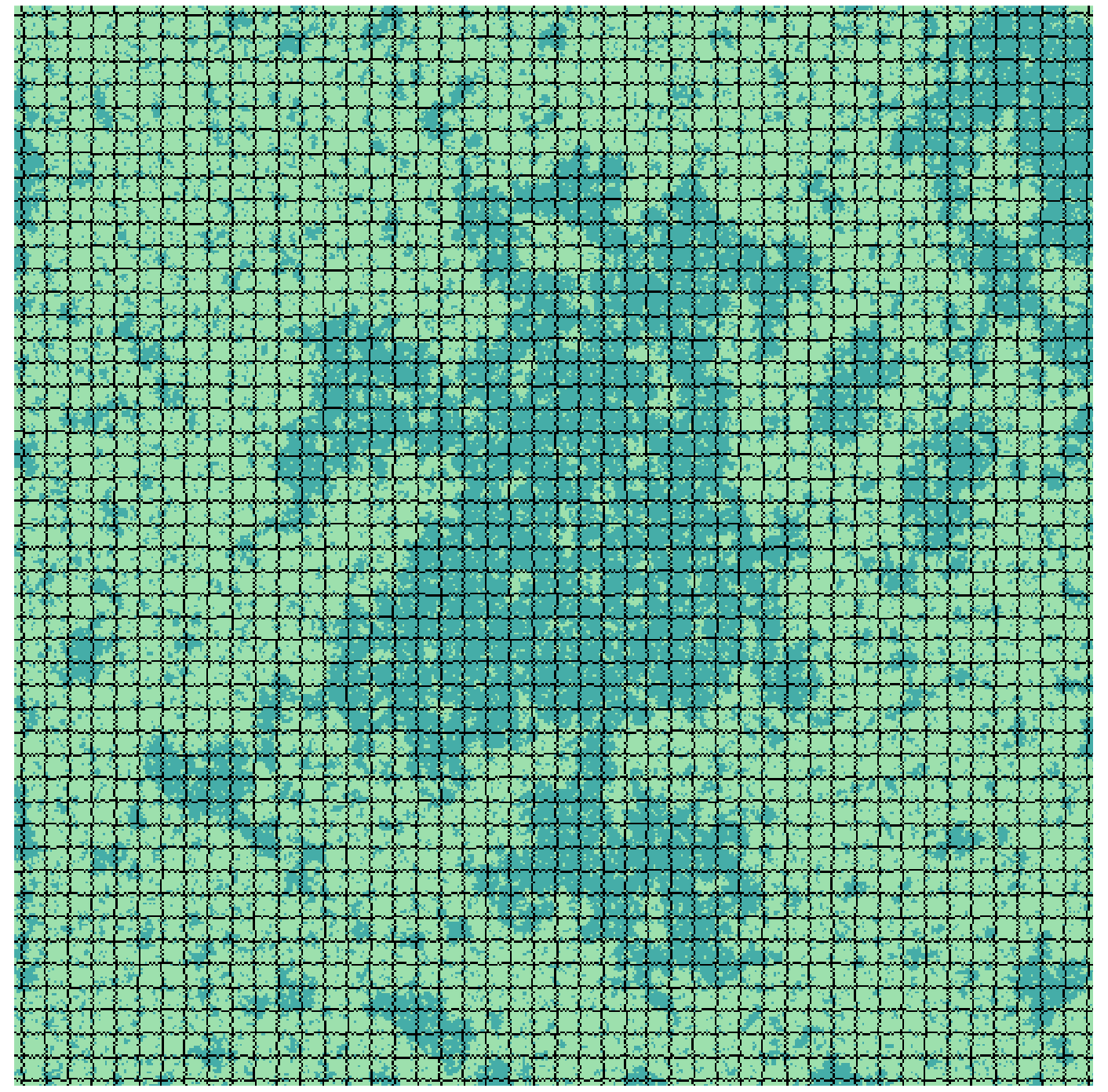}
\ig[width=.32\lw]{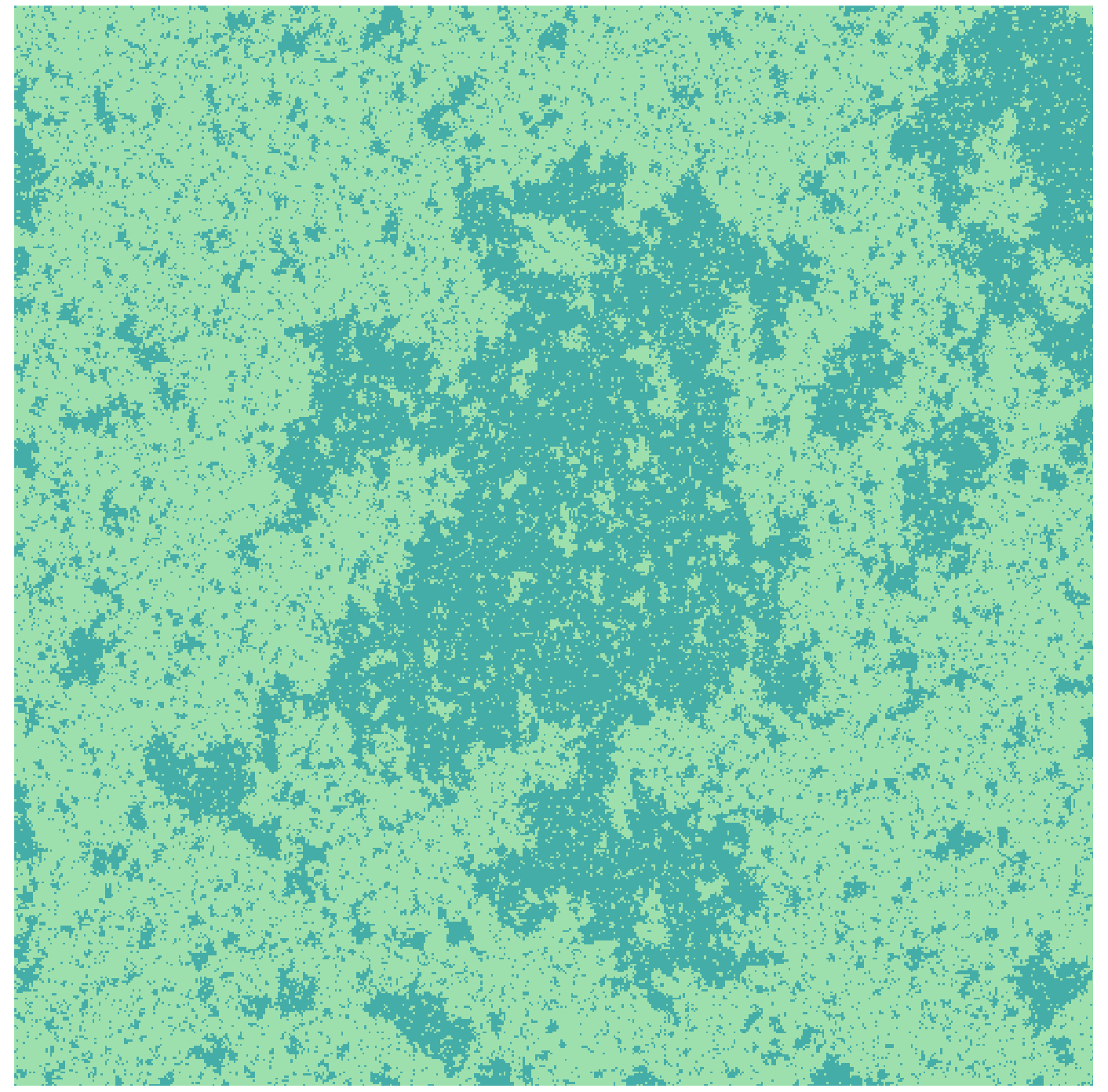}
\caption{\small{A similar chronology as in Figure \ref{FigSmallLattDil}. From the first to the second graph we apply a $\lambda = 1.1$ dilation on a 512x512 lattice. The second graph shows the patterned distribution of the holes: they are distributed along a noisy square mesh, the noise coming from the random pick of a target spin when sharing its antecedent. The presence of these patterns motivated us to introduce the random $\delta$ parameter: for different values of this parameter the mesh of holes will be slightly displaced horizontally and vertically, this way uniformizing the distribution of holes on the boundary when averaging over $\Psi_{\lambda \cdot C}$ and restoring an ``average translation invariance''.
}}
\label{FigLargeLattDil}
\ec\efgr
In the continuous limit this (discrete) operation converges naturally to the dilation we are familiar with in CFT.

On a discrete domain there is always more information on $C$ than on $\lambda^{-1} C$ and some spins may share the same pre-image \eqref{DiscreteDilFormula}. We insist on not adding spurious information, and impose a one-to-one rule: only one of them, picked at random after prioritizing boundary spins, will inherit the spin value localized at \eqref{DiscreteDilFormula}. Obviously this step will leave holes or unassigned spins. In order to fill-in the holes, we apply a heat-bath assignation: holes are filled-in according to the Ising measure induced by their neighbouring spins. This prescription is meant to restore the first neighbour correlations close to the average bulk value, see Fig. \ref{FigSmallLattDil} for a step by step graphical explanation and some additional details.

\item The second step of the cycle is a re-thermalization. As the system  is not precisely at criticality (which is impossible on a discrete lattice), the above dilation, which can be seen as an RG transformation towards the UV, took it further away. In the re-thermalization, we thus apply Monte Carlo evolution steps with the aim to bring the sample closer to criticality. This re-thermalization propagates the information contained in the new, dilated boundary condition towards the inside of the lattice. Our choice of evolution algorithm here is that of fixed-boundary Swendsen-Wang (SW) flips, see \hyperref[sec:appendixSW]{Appendix \ref{sec:appendixSW}} for implementation details and motivations.

\item Finally, we repeat the last two steps for as long as desired. As we show in the next subsection, three cycles seem to be enough to generate the correct bulk spin correlations.
\end{itemize}

We note that criticality of the initial configuration, obtained at the second point in the above procedure, is not essential for the procedure to work. Indeed, if, before the $\lambda$-dilation, the sublattice exhibited long-distance correlations up to distances of order $\xi_i$, these correlations are stretched by the blow-up steps up to orders $\xi_{i+1} \sim \lambda\,\xi_i$ (and, as mentioned, the re-thermalization steps are meant to recover the correct short-distance correlations lost by the introduction of holes). From a massive QFT point of view, after a large number of iterations, the mass gap $m\sim \xi^{-1}$ has been closed. Indeed, the philosophy of the proposed Markov chain is to aim at scaling down or reducing any energy scale, such as the mass gap $m$ or the inverse effective length scale of the whole system, in order to reach a region where the theory exhibits full scale invariance. It has been proven that in two dimensions scale invariance gets promoted to conformal invariance \cite{scaleToConf1,scaleToConf2}. In the next section, after introducing local field observables, numerical checks of the conformal invariance of the sample will be presented.

\subsection{Quality of bulk correlations}\label{ssectquality}

\begin{figure}
\bc\ig[width=.75\lw]{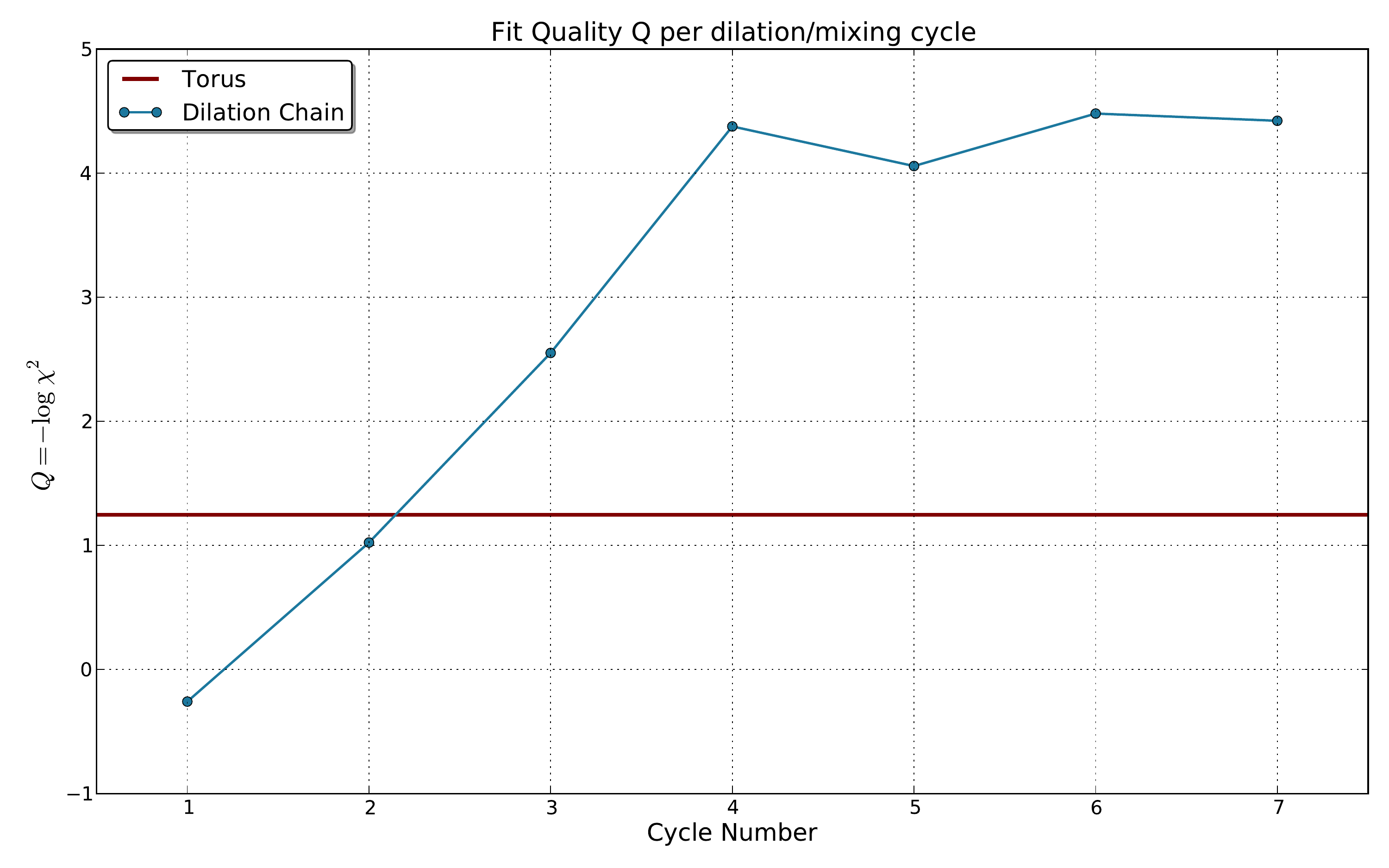}\ec
\caption{\small{The quality $Q$ as a function of the number of dilation-re-thermalization cycles, starting with a critical distribution on the torus. We choose a sublattice of size 512x512 and a dilation parameter $\lambda=2$. We fit the spin-spin correlation $\bra\,\si_{\bf i}\si_{\bf j}\,\ket$ for every $\bf i$ in the central 300x300 domain, and $\bf j$ running in neighbours up to a distance of $50$. 300 SW lattice flips are performed for every re-thermalization step, while 400 Wolff cluster flips are performed with the initial periodic boundary conditions. The correlations are obtained by averaging over the results of the SW lattice flips, omitting the first 75 flips, which are necessary in order to re-establish short-range correlations after dilations. The quality of the fit shows a significant increase during the first three dilations after which it seems to have reach its final-thermalized value. The value of $Q$ for a power law fit of the same correlation on a torus of the same size has been added as a comparison reference. It is obvious that the quality, e.g. the bulkiness, of the correlations in our chain is superior by two orders of magnitude to what can be reached by using periodic boundary conditions.}}
\label{FigQ}
\end{figure}

As a measure of the quality of the sublattice marginal, we propose a quantity which addresses the question as to if bulk correlations have been established. Consider the statistics of the product $\si_{\bf i}\si_{\bf j}$ for positions $\bf i$ and $\bf j$ in some central subdomain of the sublattice. In any sublattice marginal of an infinite system, this statistics will generate, at large enough distances, an exact power-law correlation. The spin-spin correlator $\bra \,\si_{\bf i} \si_{\bf j} \,\ket $ in the planar Ising model at criticality is well known to display the power law profile
\beq\label{Isingpower}
	\bra \,\si_{\bf i} \si_{\bf j} \,\ket \propto \frac{1}{|{\bf i}-{\bf j}|^{\frac{1}{4}}}
\eeq
at large enough $|{\bf i}-{\bf j}|$. We may then do a least-square linear fit of the set of points obtained by plotting $\log(\bra\,\si_{\bf i}\si_{\bf j}\,\ket)$ with respect to $\log(|{\bf i}-{\bf j}|)$. Two important data emerge: the fitted power $p$, and the least-square uncertainty $\chi^2$. The quality of the power law is measured by the smallness of $\chi^2$. Contributions to $\chi^2$ include both statistical noise, as well as spurious correlations leading to systematic departures from a power law. Suprious correlations may be either of universal nature, such as those induced by a conformal boundary or a periodic boundary condition, or non-universal, such as microscopic effects. For the spin variable $\sigma_{\bf i}$, microscopic effects are minimal, and the large number of samples mean that statistical noise is also small. Hence we propose the quantity
\[
	Q = -\log\chi^2
\]
as a simple measure of the quality of the bulk-marginal universal correlations. A high number represents a good quality. See Fig. \ref{FigQ} for the numerical results on the value $Q$. The increase in quality shows a decrease of power-law fitting uncertainty by a factor of about 10. After three cycles, an optimal quality seems to have been reached. Comparison with the periodic-boundary case indicates that indeed $\chi^2$ is dominated by universal correlations, and not by noise or microscopic effects.

\begin{figure}
\bc
\ig[width=.75\lw]{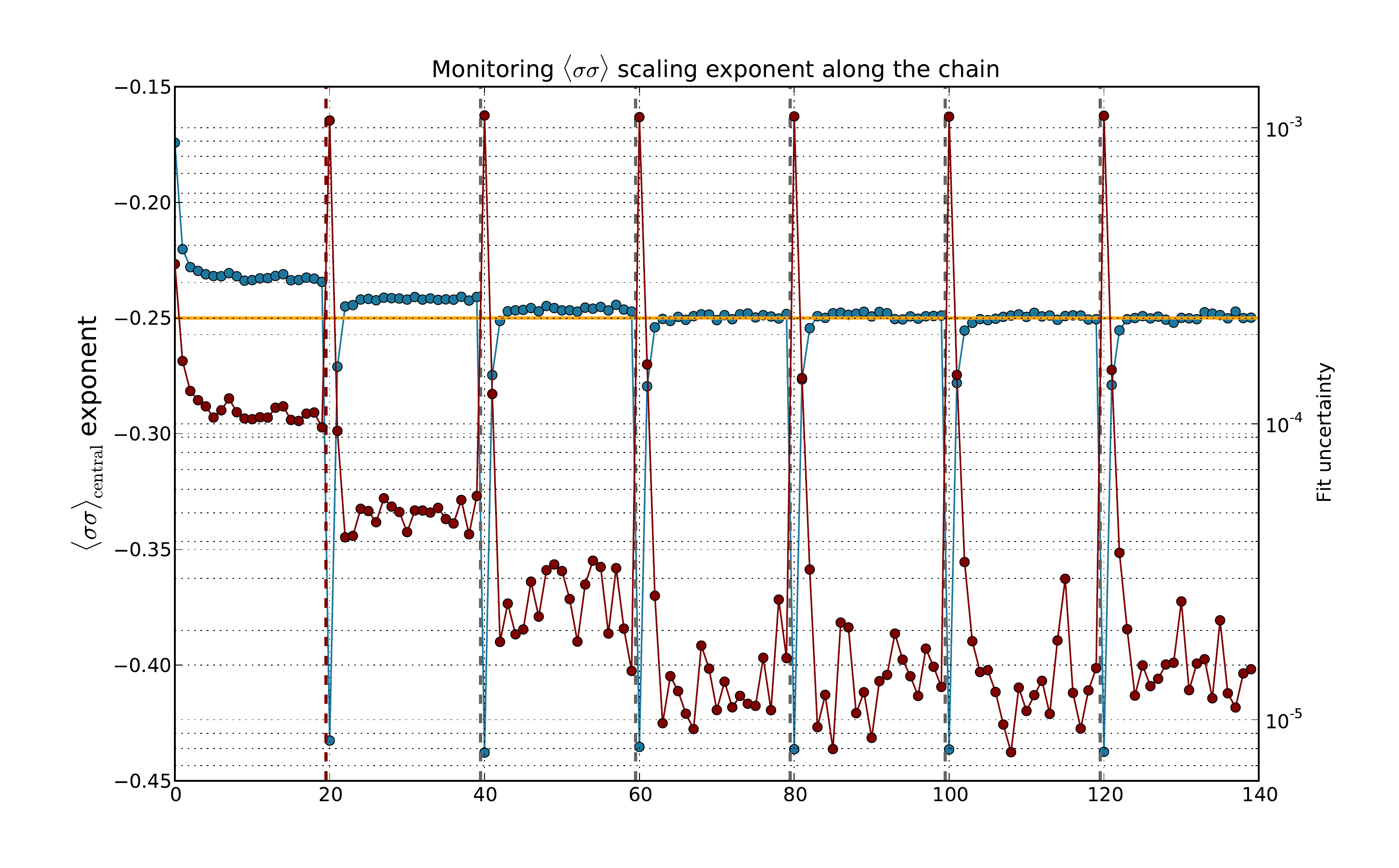}
\ec
\caption{{\small Fitted power-law and fit uncertainty for 6 dilation-re-thermalization cycles. The procedure is as in Fig. \ref{FigQ}, and here the $x$ axis is the full timeline for this procedure, by steps of 15 Wolff cluster flips or SW flips. The fit data is extracted from the set of values of $\log(\si_{\bf i}\si_{\bf j})$ with respect to $\log|{\bf i} - {\bf j}|$ for every configuration (not averaged over configurations). On the y-axis two set of data are displayed: we are superposing the graphs of the fitted value of the scaling exponent (blue dotted line) and of the uncertainty of the fit (red line). The orange horizontal line is the exponent expected in the bulk. In this graph, points lying above this line imply a slower decay of the correlations and thus lattices with an excess of spin correlations while points below it point to a lack of correlations.}}
\label{Figspinspinquality}
\end{figure}

More precise step-by-step  numerical results can be visualized in Fig. \ref{Figspinspinquality}, where the power of the power law and fitting uncertainty are displayed along the timeline. The power measured for the torus indicates a slower decay of correlations on the torus (it is clear that the torus has an excess of correlations over larger distances, coming from information going around it), but after three cycles it agrees very well with the expected bulk value. We observe that each dilation shows an abrupt jump in the direction of a lack of correlations, as the holes introduced are at best correlated to only first neighbours and thus noisy from a longer distance point of view.

Two remarks are in order. First, although the power law behaviour \eqref{Isingpower} of the correlations is expected to be true for large separations $|{\bf i}-{\bf j}|$, while the short distance behaviour would be predicted to be dominated by microscopic effects, we will see later that the spin operator - the most local operator on the lattice - shows essentially no microscopic effects in its correlation. This justifies why we measure its correlations starting at distance 1. Second, we are restricting ourselves to observations on the center of the sublattice. As we will see later and as we have argued above, the borders show some departure from the bulk expectations, which we will attempt to characterize.

As a second check of the quality of the bulk configurations obtained, we consider the average $\bra\mathcal{H}\ket$ of the energy density, defined as:
$$\mathcal{H} = -\frac{\beta }{N} \sum \limits_{({\bf i}, {\bf j}) \in {\cal E}} \sigma_{\bf i} \sigma_{\bf j}$$
where $N$ is the number of lattice sites.
To the best of our knowledge, the expected value of $\mathcal{H} $ is not known in the bulk planar triangular lattice. We may obtain an approximate value by measuring it on the torus and extrapolating ($\bra\mathcal{H}\ket_{L\times L \;\text{torus}} - \bra\mathcal{H}\ket_ {\text{bulk}} \propto \frac{1}{L}$), giving us:
\beq\label{valeps}
\bra{\mathcal{H}}\ket_{\rm bulk} \approx -0.54936\,(4)
\eeq

\begin{figure}
\bc
\ig[width=.8\lw]{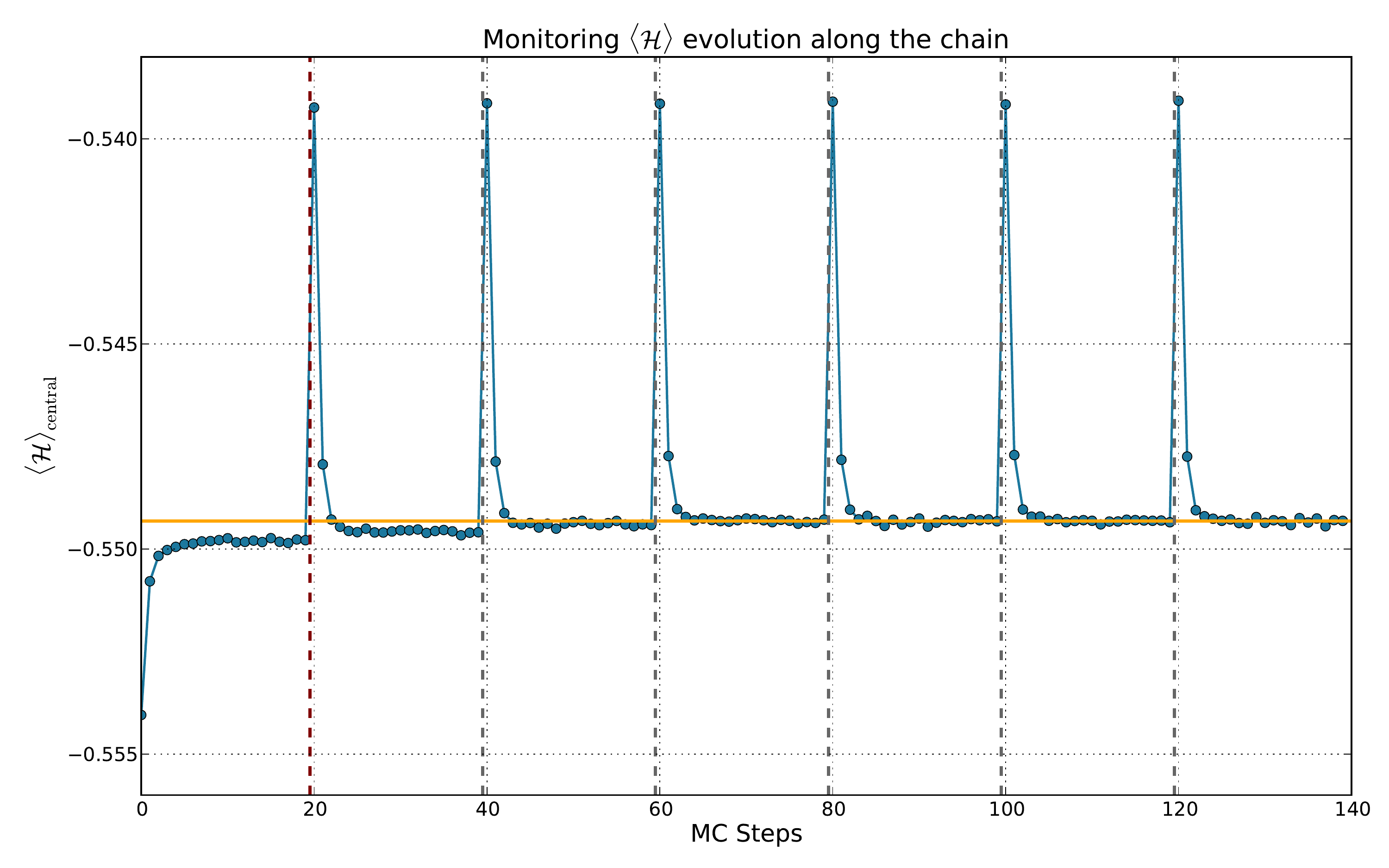}
\ec
\caption{{\small Along the x-axis is the timeline of the evolution history we described as in Fig. \ref{Figspinspinquality}. Before the red dashed vertical line are the torus Wolff steps. As we see, and as is well known, the value converges to a constant departed from the bulk (orange horizontal line). After the red vertical line, the cycle of dilations and lattice flips starts.}}
\label{Figepsilon}
\end{figure}

The numerical results are displayed in Fig. \ref{Figepsilon}. We clearly see, again, that the dilation induces a jump in the value of the energy density: even though we are using a heat-bath assignation method, there are holes that happen to be next to each other (recall that here $\lambda = 2$ so that at least $\frac{3}{4}$ of the sites post dilations are unassigned). The jump happens in the upward direction, because of the decrease of $\bra \si_{\bf i} \si_{\bf i+1} \ket$ (here $\bf 1$ represents any vector pointing to a nearest neighbour), which is a loss in next-neighbour correlations. Despite that jump, the evolution very quickly stabilizes, and after three dilations we can already see a convergence to a thermalized value of $\bra {\mathcal H} \ket$ sitting on the critical value \eqref{valeps}.

It is interesting to remark here that in both spin-spin correlations and energy density average, the thermalization seems to happen passed the third dilation-re-thermalization step: as if first neighbour ($\mathcal H$) and further neighbours ($\si_{\bf i} \si_{j}$) thermalize almost simultaneously. Both sets of data show how dilations and lattice flips work hand in hand in the algorithm in order to wash away the initial boundary information and to bring the sublattice distribution closer to a marginal of the planar Ising model.

We have not observed any significant autocorrelations between realizations separated by a single dilation-re-thermalization cycle. This indicates some effectiveness of the Markov chain at sampling the indenpendent bulk marginals; although autocorrelations are likely to appear as $\lambda \to 1$.

Different runs were also made for other values of $\lambda \in (\frac{10}{9},2)$. The observed convergence is qualitatively unchanged, up to the general rule that the number of dilations needed to achieve mixing to a bulk marginal increases as lambda decreases (no trivial formula seems to describe this relation).

The same Markov chain has been implemented on a square Ising lattice at its critical coupling. The results on the monitoring of the convergence to a bulk marginal are very similar. This indicates that, correctly implemented, the map $R_{\lambda}[\cdot]$ is a universal method to sample finite domains of a critical lattice system.

Finally, we remark that many ``arbitrary'' choices were made in our implementation of the Markov chain, including that of the starting point, of the hole-filling prescription and of the evolution algorithm. These choices are the result of many trials and errors and were motivated by the criteria of computational speed and quality of sampling.

\subsection{Remnant boundary effects}
\label{sec:deltaEexponentsV2}

\begin{wrapfigure}{l}{0.45\textwidth}
\centering
\ig[width=.65\lw]{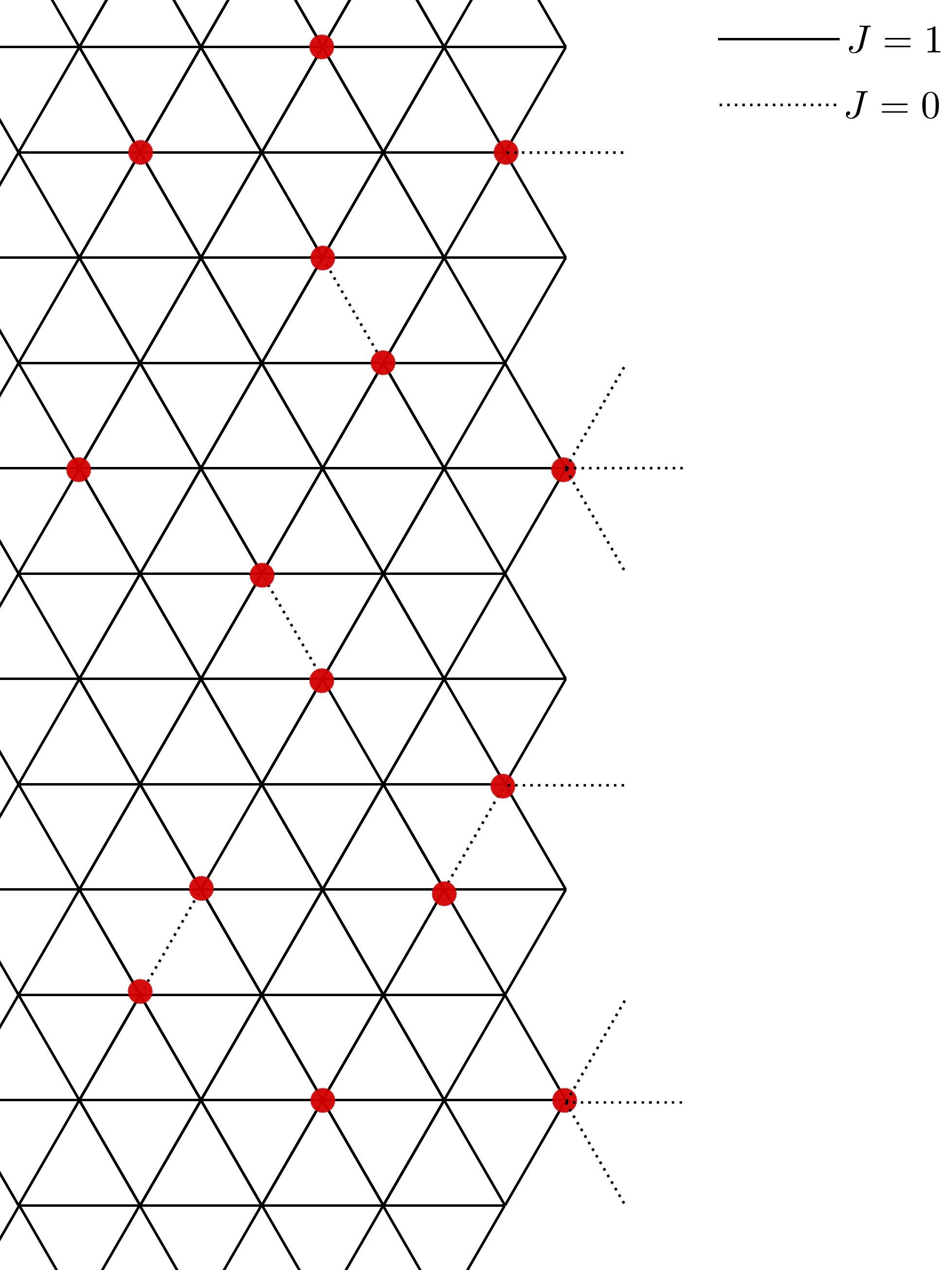}
\caption{\small{Distribution of holes - marked as red points - near the border of the triangular lattice. In the local heat-bath prescription, some bonds - marked by dotted lines - in the measure $\exp[\beta \sum_{({\bf i},{\bf j})\in{\cal E}} J_{{\bf i}{\bf j}}\sigma_{\bf i}\sigma_{\bf j}]$ are missing, $J_{{\bf i}{\bf j}}=0$.}}
\label{MissingBondsBoundaryFig}
\end{wrapfigure}

Residual effects from holes being assigned on the boundary cannot be fully corrected by any prescription or re-thermalization. There are thus remnant boundary effects, which it is important to attempt to characterize.

Conformal boundaries (infinite magnetic field or infinite temperature applied on the boundary spins) or periodic boundaries are known to induce a departure of the average energy density, decaying as a power law (with decay exponent 1) in the distance to the nearest boundary or as a power law in the lattice size respectively. Our data shows a behaviour similar to these ``canonical setups'' with the boundary condition $R_\lambda^N[C]$ (after the $N^{\rm th}$ step) inducing a power law on the expectation value of the lattice operator $$\varep_{\bf i} = \sum_{{\bf j}\, :\,\bra{\bf i},{\bf j}\ket \in {\cal E}} \sigma_{\bf i} \sigma_{\bf j} + \frc{2}{\beta_c} \bra{\mathcal H}\ket.$$ It is detailed in (\ref{defener}) as the fluctutation of the lattice energy operator, which renormalizes into the energy operator $\phi_{1,3}$ from the Ising Kac table. On a planar lattice we would expect $\bra \varep_{\bf i} \ket = 0$.

Numerically, see Fig. \ref{EvevFig}, we observe that this induces an expectation value on the lattice energy fluctuation field $\bra \varep_{\bf i} \ket \neq 0$. This is expected to be a consequence of the unthermalized spins on the boundary (the number of which is proportional to the dilation parameter $\lambda$). Indeed, recall that the blow-up procedure included a prescription for holes left within the region $C$ (after blow-up from $\lambda^{-1} C$). The prescription used is to locally thermalize using the information of the assigned neighbours. Links to unassigned neighbours (holes) are effectively set to 0. Holes are thus assigned random locally-thermalized values, but with a modified Hamiltonian, corresponding to the introduction of link defects (see Fig. \ref{MissingBondsBoundaryFig}). In the interior of $C$, this is of little consequence because the re-thermalization step guarantees that short-distance correlations are correctly re-constructed and these defects are washed away. However, at the boundary, these defects remain. Since interior re-thermalization is performed with fixed boundaries, these defects are expected to have an effect on interior distribution.

 The profile we measure for this expectation value is a decaying power law in the distance to the closest border. Interestingly, the decay exponent seems to be dependent on the latest dilation parameter, and to take a continuum of values. For the values of $\lambda$ taken, the exponent always lies between 1 and 2. Since it is larger than 1, the boundary effects are smaller than those coming from a conformal boundary condition. The decay exponent increases, and the amplitude decreases, as $\lambda$ is brought nearer to 1,  and these effects can be made to be very small at distances greater than about 20 sites from the boundary.

It has been observed in different setups \cite{defectSpectrum1,defectSpectrum2,defectSpectrum3} that the critical Ising model with a line of defects shows a continuous spectrum of perturbations depending on the defect strength. This defect strength can be a departure from the critical coupling such as in \cite{defectSpectrum1}. In this light, it is tempting to identify our observations with this phenomenon: the line of defects can be interpreted as the sublattice boundary and the density of cut bond as a defect strength, an effective departure from the critical coupling.

We also provide in Appendix \ref{appCFTmarkov} a basic CFT analysis of the effects of the lattice on the Markov chain. This analysis is expected to give only a maximal value for the decay exponent of the boundary effect, predicted there to be 2 (in agreement with the observed values, all lesser than 2).

\bfgr
\bc
\ig[width=.8\lw]{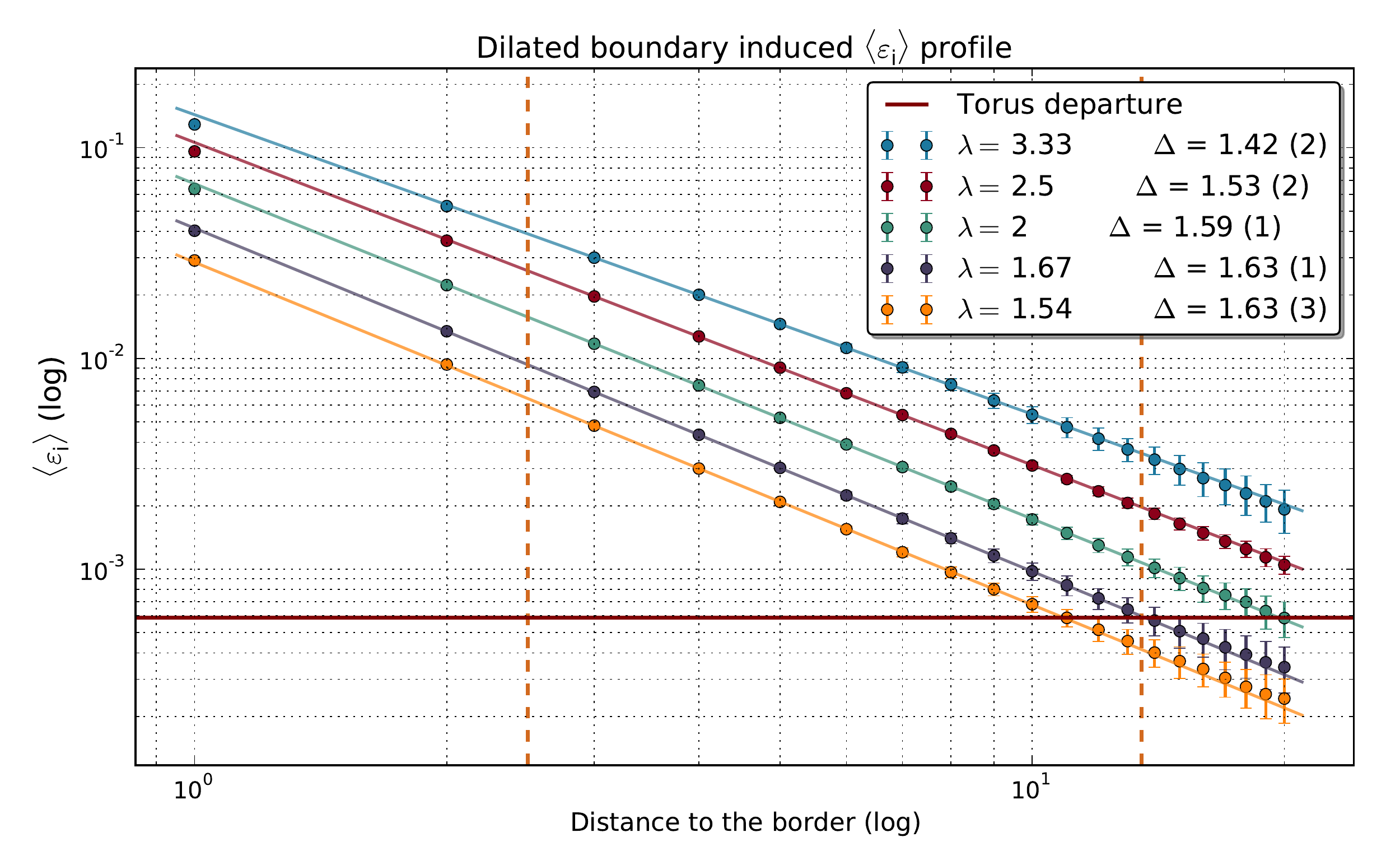}
\caption{\small{The departure of the average energy density as a function of the distance to the closest boundary (log scaled $x$-axis) for different dilation factors $\lambda$. The power law decay is apparent on this graph and exponent fits were performed in the range bounded by the two dotted yellow vertical lines. Very interestingly, not just the offset of the departure is dependent on the fraction of holes on the boundary -- the last dilation parameter $\lambda$ -- but the fitted exponent as well. For each value of $\lambda$, we fit an exponent $\Delta$ which lies in a range from 1 to 2. At the critical temperature these boundary effects are carried away with infinite range, thus over the whole lattice. The red line was added to show the average departure $\delta\varep = \big|\bra \varep^{\rm torus}_{L{\rm x}L} \ket + \frac{2}{\beta_c} \bra\mathcal{H}\ket\big| \propto L^{-1}$ found on a same-size torus. This tells us that the boundary effect, although carried over an infinite range, is, in magnitude, insignificant in comparison to the effect on the torus, as soon as observation is made more than 20 lattice units away from the border and with dilation factors $\lambda \leq 2$.}}
\label{EvevFig}
\ec
\efgr

\section{Numerical verifications}\label{sectnumer}

\subsection{Fractal dimensions of the loops}

At criticality, the system is scale invariant and it is known that the clusters of equal spins and their boundaries form fractal sets \cite{nienhuis}. The values of the fractal dimensions are well known from Coulomb Gas techniques or more recently derived from SLE arguments \cite{beffara}. The possibility of measuring the effective fractal dimension of the loops' boundaries and of the domains lying between the loops offer a good way of verifying that the samples produced from the proposed Markov Chain are indeed critical Ising samples. We emphasize that these are \textit{effective} fractal dimensions, as objects defined on a finite lattice cannot be rigorously fractal; nevertheless they still very nearly exhibit, for instance, fractional scaling behaviours.

The sample used for the estimations below were taken in a pool of size $\sim 60\,000$ sublattices of size 512x512. Each was generated by the Markov chain detailed in section \ref{sectmarkov}, stored after each completion of the re-thermalization steps. The dilation parameter used along the chain was $\lambda = \frac{5}{4}$. The two fractal dimensions -- of the cluster boundaries and of the clusters themselves -- were checked independently employing two different methods. 

\subsubsection*{Box counting}

\vbox{
\begin{wrapfigure}{l}{0.48\textwidth}
  \centering
    \includegraphics[width=0.48\textwidth]{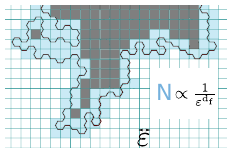}

  \caption{\small{Illustration of the box counting method applied to a fraction of a large loop. The number of intersections of the loop with the mesh scales with the mesh size $\varep$ as a power law with the (opposite) fractal dimension as exponent.
If we add to the count the grey boxes inside the loop (excluding domains inside possible inner loops), the scaling of this quantity will give the ``mass'' exponent of the cluster.
}}
\label{fractalDim}

\end{wrapfigure}

Box counting is an effective method for computing the fractal dimensions of any object by measuring the scaling exponent of the number of its intersections with a mesh of varying size. For sufficiently small meshes, it is expected the the number of intersections scale as $N(\varep) \propto \varep^{-d_f}$, where $\varep$ is the mesh size and $d_f$ the fractal dimension of the geometric object (see Fig. \ref{fractalDim}).
 In the sampled sublattices, looking only at the loops with diameter (defined as the maximum distance between two polygon edges) greater or equal to 40 lattice units, we find the average value of the fitted exponents to be $1.375 \pm 0.05$. It is known from $\rm SLE_{\kappa}$ that the fractal dimension of Ising is given by ${\rm min}(1+\frac{\kappa}{8},2)$ with $\kappa_{\rm Ising} = 3$. Our estimation is well centered on the theoretically known value although with a quite large uncertainty.

By the same algorithm, we estimate the fractal dimension of the cluster domains to be $1.95 \pm 0.02$, to be compared with the theoretically predicted value: $\frac{187}{96} = 1.948\dots$. Here we selected clusters whose exterior bounding loop is of diameter at least 30. Again, our estimation suffers from a relatively large fit uncertainty. The culprit seems to be in the fact that box counting method offers a rather poor power-law aspect, with important statistical deviations due to the finiteness of the mesh size.
}

\subsubsection*{Bulk finite-size scaling}

The same exponents have long been evaluated on the torus by the use of finite-size scaling. This is based on the assumption that the mean loop-length and cluster-mass of the longest loop or heaviest cluster, respectively, scales as a power law in the periodic lattice size $L$, with exponent the associated fractal dimension, at least at leading order \cite{saberi}.

Focussing on the fractal dimension of the length of loops, on a torus of size $L$x$L$, we have the following:
$$\bra { \underset{\mathscr{L}}{\rm max}\; {\rm length}(\mathscr{L})} \ket_{L{\rm x}L}\ \sim \ L^{d_f}\  +\ \text{sub-leading terms} $$
where $\rm max$ runs over the set of closed loops $\mathscr L$ in each toroidal lattice.

To our knowledge no similar estimator is known inside the bulk. By scale invariance arguments, we should expect that an estimator looking for the longest loops on a finite $L$x$L$ sublattice of the infinite planar lattice should display the same power law at leading order, as the fractal dimension is not dependent on the boundary conditions. However, as we measure only the loops entirely inside the subsection, for small values of L, the impact of excluding the loops touching the borders becomes significant. Our estimation for $L \in [100,1000]$ was unsatisfying. To counter this issue, we decided to refine the estimator by introducing an arbitrary parameter $\alpha > 1$. The recipe is the following: in order to avoid the small boundary effects of the bulk marginal, we cut a subsection $S$ inside a larger bulk marginal of size $L{\rm x}L$. The subsection $S$ is of size $\alpha L' {\rm x} \alpha L'$ for $\alpha L'<L$ and $\alpha>1$, and in it, we run over all $L'{\rm x}L'$ boxes. The estimator will look at the longest loop fitting in the running $L'{\rm x}L'$ box inside $S$. For a fixed value of $\alpha$, we vary $L'$ and expect to have an average maximal length scaling as $(L')^{\rm d_f}$.

This new estimator, named $M(L',\alpha)$, should factorize into functions of $L'$ and $\alpha$ separately.
Indeed, under a scale transformation $L' \rightarrow b L',\; \alpha \rightarrow \alpha$, we should have:
$$M(bL',\alpha) = b^{\rm d_f} M(L',\alpha), \qquad \forall b > 0$$
constraining:
$$M(L',\alpha) = L'^{\rm d_f} M(1,\alpha).$$

The evaluation of this new estimator for the length and mass exponents -- on the same samples on which we ran the box counting estimation -- using a value $\alpha =5$ gives Fig. \ref{BulkFSSfig}.

\bfgr
\bc
\ig[width=.8\lw]{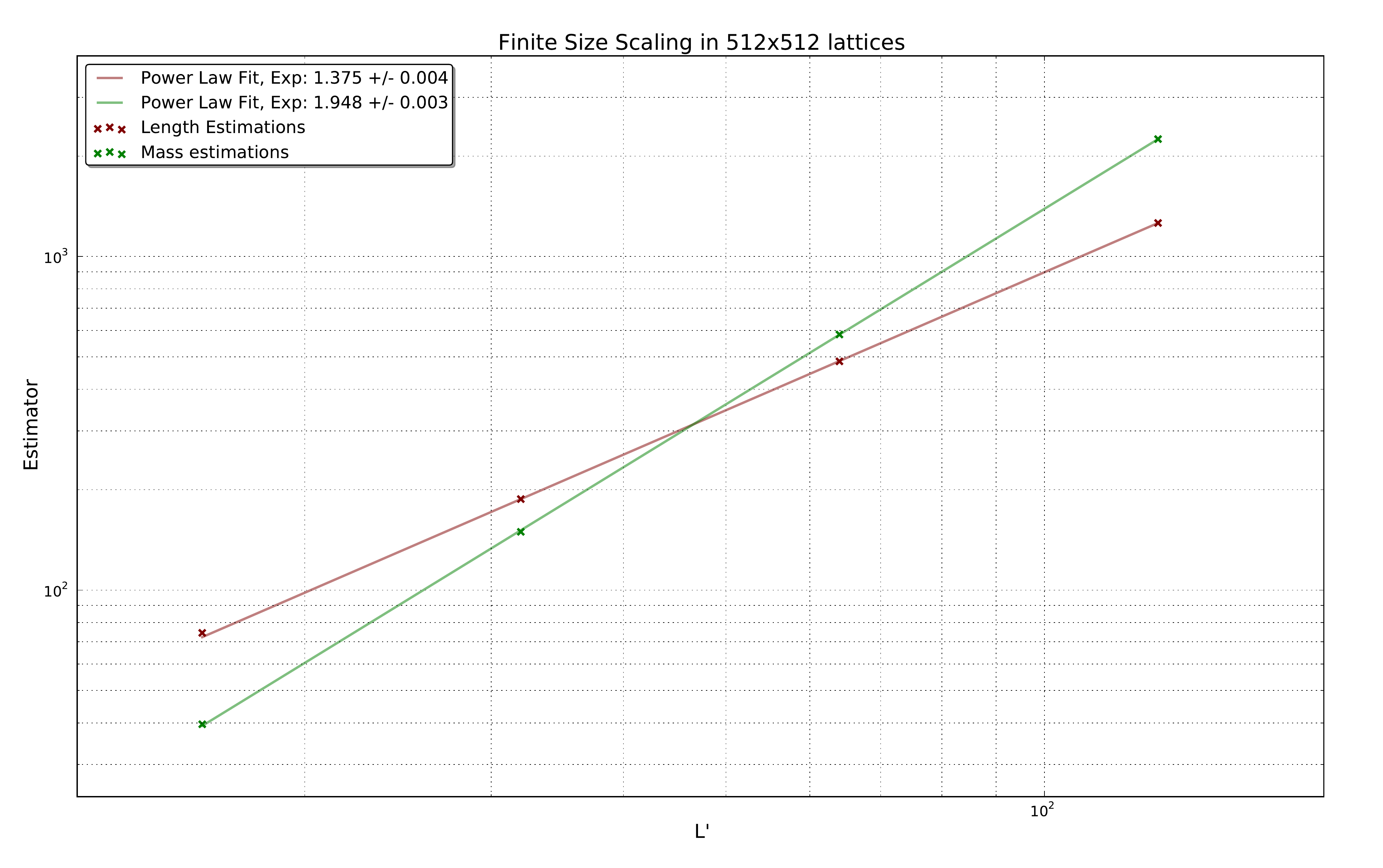}
\caption{\small{Log-log plot of the average maximum length/mass of a cluster (y axis) fitting in square domain $L'$x$L'$ ($x$ axis) and their fit. The power law behaviour is indisputable. Uncertainties were added but smaller than the points width.}}
\label{BulkFSSfig}
\ec
\efgr

Just as for estimations using box counting, we find agreement with the theoretically known value for the fractal dimension of the boundaries, the progress is in the reduction of its uncertainty by one order of magnitude. Regarding the mass exponent, we find here better agreement with a fitted value of $1.9476 \pm 0.0029$.  This is a significant improvement from the box counting estimation, and, to our appreciation, sufficient on its own to convince of the relevance of adapting finite-size scaling estimators to bulk subsections. Here as well, we gained an order of magnitude on the precision. It is interesting to stress that this implementation of finite-size scaling in the bulk seems to be matching the precision magnitude of finite-size scaling on the torus \cite{saberi}.

These results on their own they are not enough proof that our samples have features close to those expected in the bulk, as fractal dimensions are not expected to be affected by conformal boundaries. They are only verifications that the samples are very close to the critical point. The next subsections provide in-depth numerical analyses of proper bulk properties: bulk correlation functions.

\subsection{Two- and three-point functions of spin and energy fields}

As emphasized in subsection \ref{ssectquality}, two-point functions are good indicators of how close to critical Ising bulk marginals the MCMC sampler gets. Here we look closer at the correlations over larger distances.

Recall that CFT scaling fields have two-point functions of the general form
$$\langle \,{\cal{O}}(z_1)\, {\cal{O}}(z_2)\, \rangle = \frac{\textrm{e}^{-2i\si_{\Or}\theta_{z_1z_2}}}{|z_{12}|^{2\Delta_{\Or}}}$$
where $\si_{\Or}$ is the spin of the field and $\Delta_{\Or}$ its scaling dimension (here $z_{12}=z_1-z_2$ and $\theta_{z_1z_2} = {\rm arg}(z_{1}-z_{2})$).

First consider the spin variable $\sigma_{\bf i}$. This scales to a CFT primary field $\sigma(z)$ as $$\sigma_{{\bf i}} \sim N_{\sigma} a^{\Delta_{\sigma}}\sigma(z)\quad (a\to0,\;z=a{\bf i} \ \mbox{fixed}).$$ The results for the absolute value of its two-point function are displayed in Fig. \ref{SScorrFig}. This was obtained with up to 200 lattice spacings in every possible direction, and averaging over approximately 20 000 samples of size 512x512, taken after the re-thermalization steps (after having achieved complete mixing) and separated by at least one dilation. Excellent agreement is found with a power law behaviour, and one can see that the results are much nearer, especially at large distances, to those expected for bulk marginals than equivalent results numerically evaluated in a torus geometry.
\bfgr
\bc
\ig[width=.8\lw]{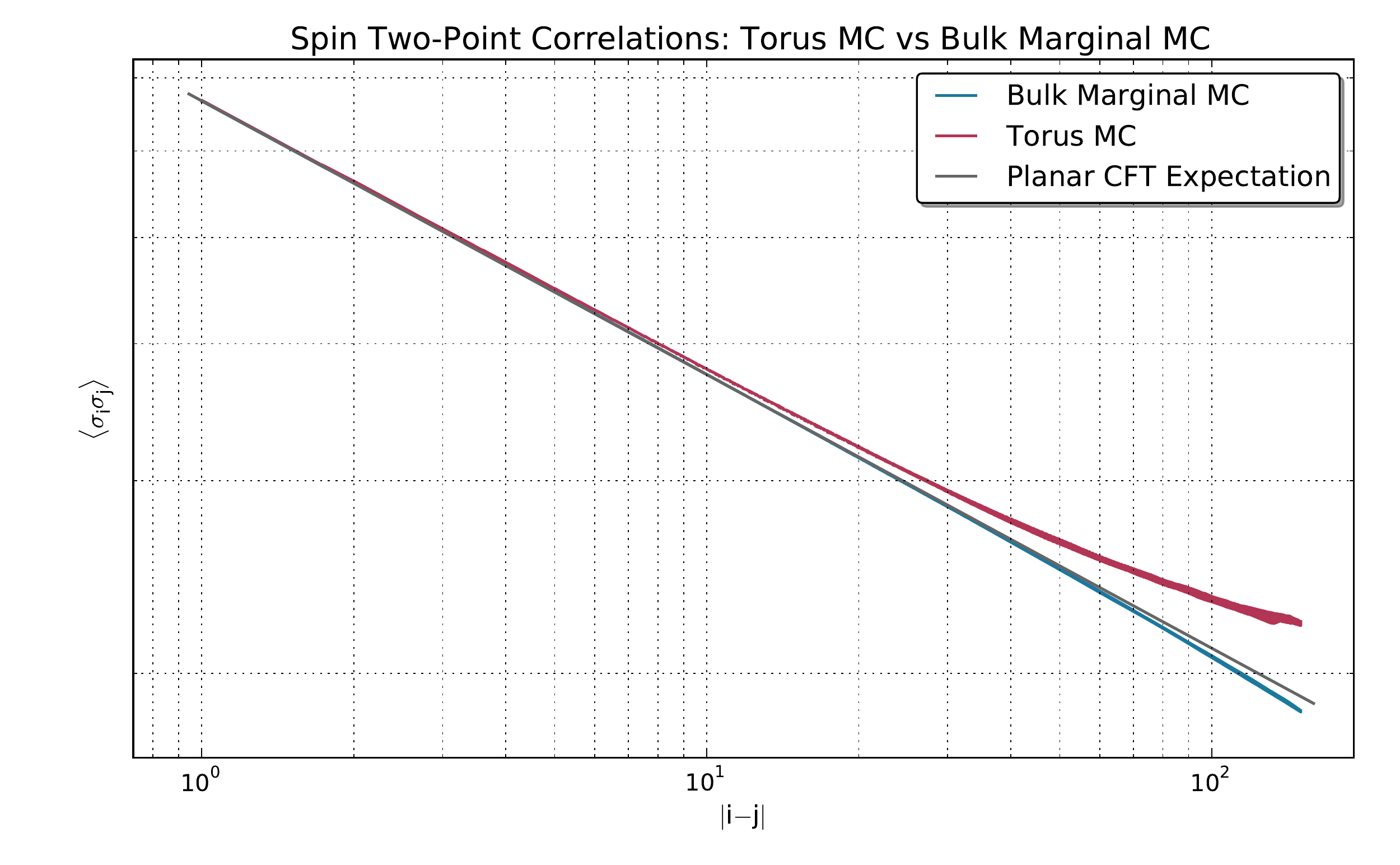}
\caption{\small{The red line corresponds to the results from the MCMC sampler. Both axes are log scaled. We added an estimation of the correlator on the torus for the same size (blue), as well the predicted value on the bulk (grey). It is clear that the results of the Markov chain sit closer to the infinite-volume analytical result, a small departure in the direction of a lack of correlations appears over larger distances. This is supported by the fitted scaling exponent on this graph: $2\Delta_{\si} = 0.24996\,(2)$. Recall from subsection \ref{ssectquality} that the ratio of the $\chi^2$ between the bulk and the torus fit is 0.019, meaning that our correlation exhibit a power law behaviour significantly more manifest, as this graph supports. We assume the lack of correlations appearing in the far right of the tail to be induced by the same boundary effects affecting the expectation value of the energy density operator we described in subsection \ref{sec:deltaEexponentsV2}.}}
\label{SScorrFig}
\ec
\efgr

We also verified bulk rotation invariance. The spin variable is spinless, $s_\sigma=0$, by definition (more precisely, it is invariant under the $\Z_6$ lattice rotation symmetry group; of course this is expected to be promoted to full $SO(2)$ rotation invariance in CFT). However, discrepancies from bulk marginals may result in residual boundary effects from the square-shaped boundary. We estimated such effects by fitting the angular dependence of the correlator onto a cosinus function in order to estimate an effective $s_\sigma$. We looked at the correlator in three different directions $\theta_{{\bf i}{\bf j}} \in \{0,\pm\frac{\pi}{3}\}$ (which are related by lattice rotations), followed by a reweighting of the initial data as to suppress the dependence on the distance $|z_{12}|^{2\Delta_{\si}}$. The fit gave the estimate $s_\sigma=4.10^{-13} \pm 2.10^{-5}$, in good agreement with a theoretical value of 0.

The energy density operator may be defined by:
\beq \label{defener}
\varepsilon_{\bf i} =\sum_{{\bf j}\, \in \,{\cal N}({\bf i})} \sigma_{\bf i} \sigma_{\bf j} + \frc{2}{\beta_c} \bra{\mathcal H}\ket
\eeq
where ${\cal N}({\bf i}) := \{{\bf j}:({\bf i},{\bf j})\in {\cal E}\}$, that is, the sum runs over the positions ${\bf j}$ that are neighbours of the site ${\bf i}$. The second term aims at suppressing the non-universal expectation value $-\frac{2}{\beta_c}\bra\mathcal{H}\ket = 3.99938\,(9)$ of the first term\footnote{The authors have not found a reference for this value in the literature on the Ising model, but are suspecting an exact value equal to 4.}, so it behaves as a CFT primary operator $\varepsilon(z)$ at large distances,
\[
	\varep_{{\bf i}}
    \sim N_{\varep} a^{\Delta_{\varep}} \varep(z)
    \quad (a\to0,\;z=a{\bf i} \ \mbox{fixed}).
\]

The energy field has spin 0 and scaling dimension 1. We analyzed the two-point function with data collected over the same samples as those discussed above. We restricted to  separations smaller or equal to 50 lattice units as beyond these, the noise was becoming dominant (it seems that the variance on the energy correlations is stronger than on the lattice spin correlations). We also excluded from the fitted data all distances smaller than 6, as they seemed to suffer from some microscopic effects. Our fit gives a scaling weight of $\Delta_{\varepsilon} = 1.0028\,(4)$, falling satisfactory close to the theoretical value of $\Delta_{\varep}^{\rm CFT} = 1$. See Fig. \ref{figee}.
\bfgr
\begin{center}
\includegraphics[width=.8\lw]{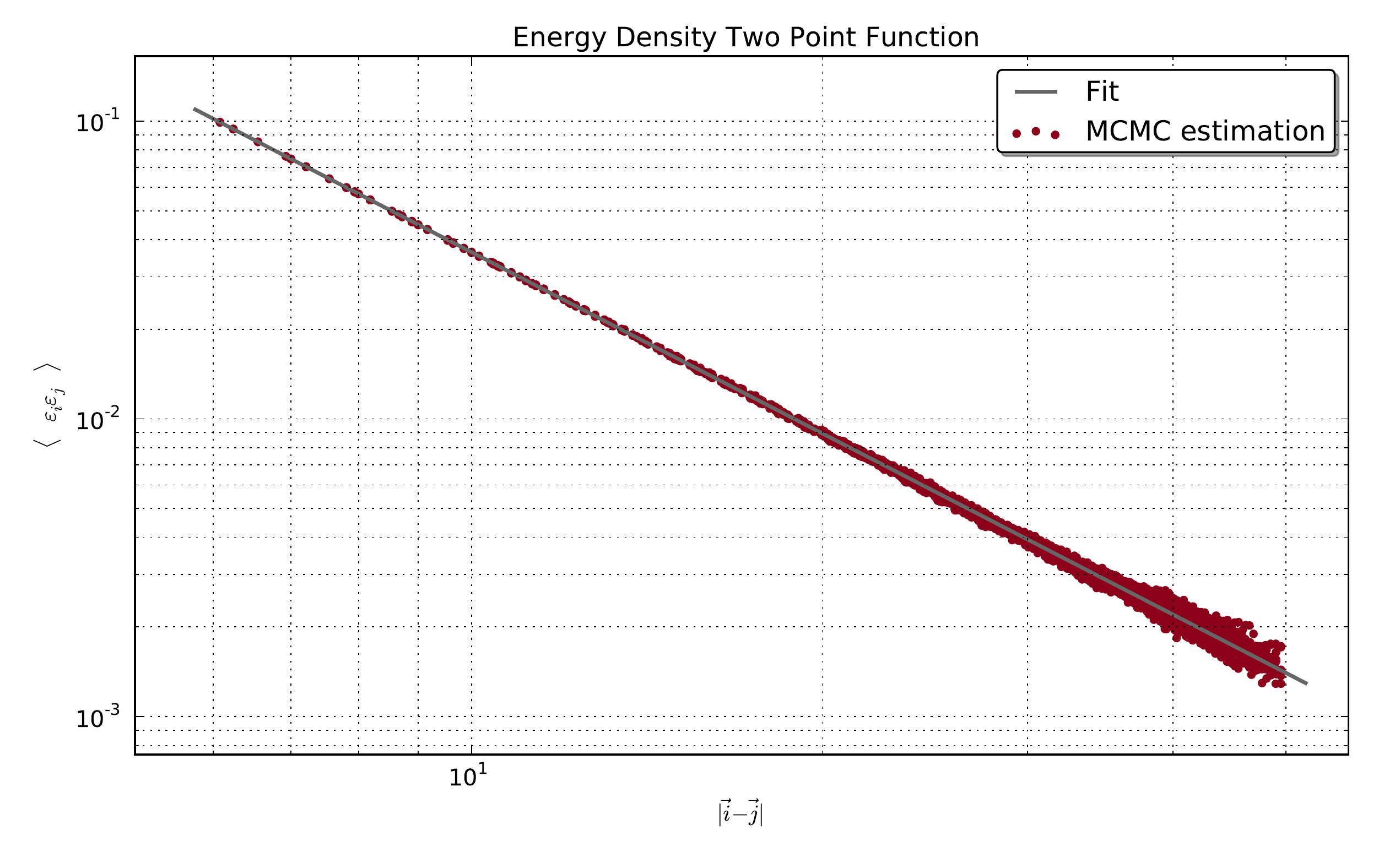}
\caption{\small{The energy-energy correlator in a log-log plot. This excludes the smallest separations -- which showed a small excess of correlation -- and the power law behaviour is clear. We substracted the disconnected part ${\mathcal H}^2$, in order to have an operator with vanishing expectation.}}
\label{figee}
\end{center}
\efgr
We also repeated the three-directional fit we had done for the lattice spin variable, and could fit an effective spin for the energy density to a value of $7.10^{-4} \pm 3.10^{-2}$; once again in good agreement with a theoretical value of 0.

From the power-law fits, the non-universal normalization factors $N_\sigma$ and $N_\varep$ are evaluated to:
\begin{eqnarray}\label{Nsi}
N_{\si} &=& 0.8166\,(4)\\
N_{\varep} &=& 1.927\,(2).
\end{eqnarray}
This allows us to estimate a first ``dynamical'' element of the CFT data (quantities that are not directly constrained by behaviours under global conformal symmetry): the structure constant $C_{\si \si\varep}$.

We evaluated the $\langle \varepsilon_{\bf i} \sigma_{\bf j} \sigma _{\bf k} \rangle$ correlator in the limit where the position ${\bf j}$ of the first spin variable is close to the position ${\bf i}$ of the energy variable. A result of an operator product expansion in CFT is:
$$\langle\, \varep_{\bf i } \sigma_{\bf j} \sigma_{\bf k} \,\rangle \stackrel{0\ll|{\bf i} - {\bf j}|\ll|{\bf j} - {\bf k}|}{\sim} \frac{C_{\sigma\sigma\varepsilon}}{|x-y|\,|y-z|^{\frac{1}{4}}}\ a^{\frc54}\,N_{\si}^2 N_{\varep}$$
with $x=a{\bf i}, y=a{\bf j}$ and $z=a{\bf k}$ fixed and where $a\to0$ is the lattice spacing.

Looking at this observable on horizontal directions in 512x512 lattices for $|{\bf i} - {\bf j}|<50<|{\bf j} - {\bf k}|$, excluding distances under 30 lattice units from the boundaries in every direction, and running over 5 000 lattice samples, gives the output displayed in Fig. \ref{figESS_SS}. There, the ratio with the two-point spin correlator has been taken, thus taking away the factor $|y-z|^{-\frc14} a^{\frc14} N_\sigma^2$. The power law behaviour is obvious in this log-log plot, from the shortest distance (that is, it seems the requirement $0\ll |{\bf i}-{\bf j}|$ is satisfied even at distances $|{\bf i}-{\bf j}|=1$) and even up to large distances. Fitting on a power law gives a scaling weight of $1.008 \,(1)$, for the remaining power of $|x-y|^{-1}$, once again in satisfactory agreement with the theoretical value.
\bfgr
\begin{center}
\includegraphics[width=.8\lw]{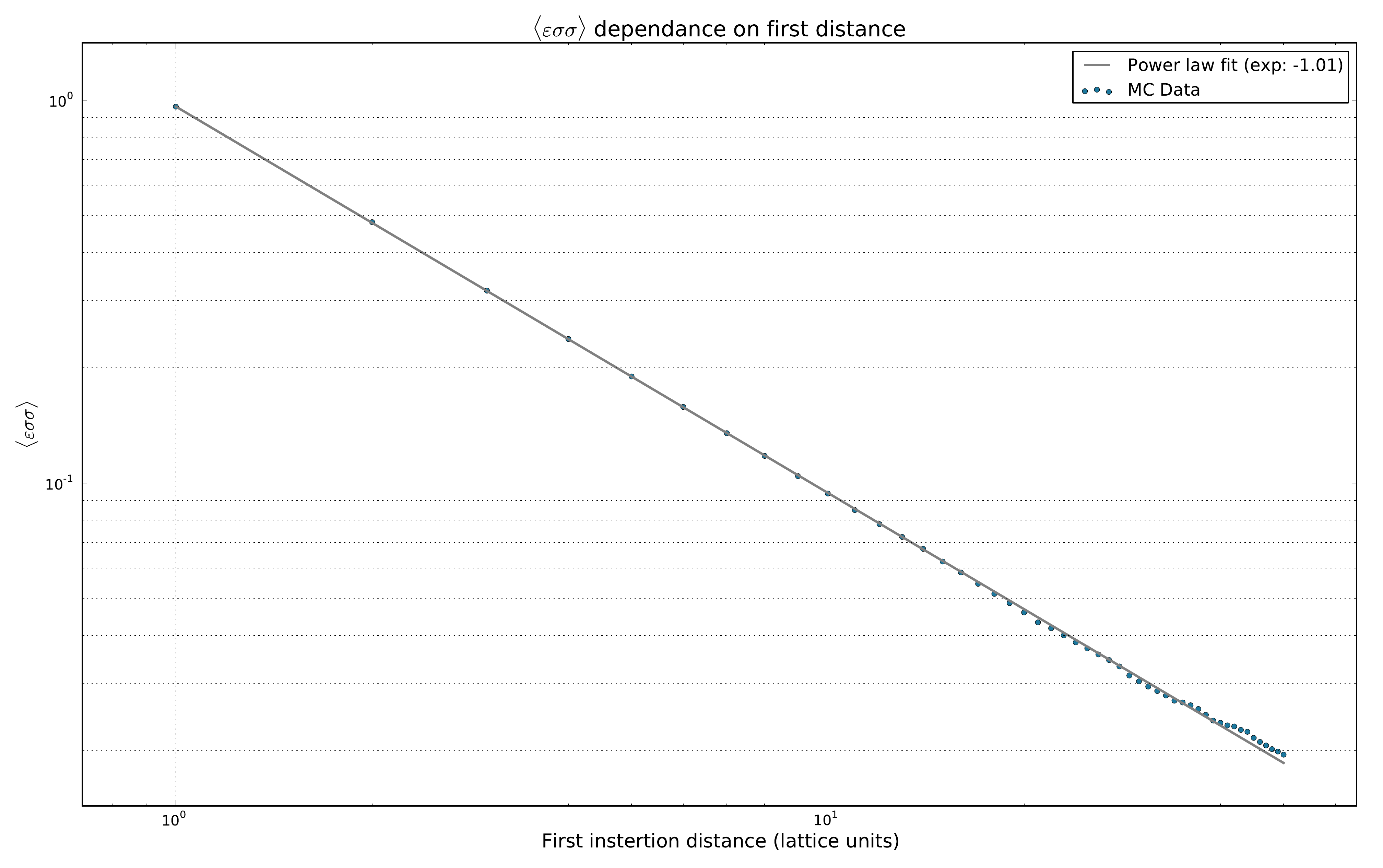}
\caption{\small{Log-log plot of the ratio $\;\frac{\langle \varepsilon_{\bf i} \sigma_{\bf j} \sigma_{\bf k} \rangle}{\bra\si_{\bf j} \si_{\bf k}\ket}$ versus $|{\bf i}-{\bf j}|$ for $|{\bf j} - {\bf k}| \ll |{\bf j}-{\bf k}|$.}}
\label{figESS_SS}
\end{center}
\efgr

The fit of the offset $C_{\si\si\varep} N_{\varep}$, with the previous estimations of $N_{\varep}$, allows us a numerical estimation of the structure constant $C_{\sigma\sigma\varepsilon} = 0.498\,(3)$, in good agreement with the theoretically known value of $\frac{1}{2}$ (see e.g. \cite{df_cft}).

This coefficient had already been derived numerically in a Monte Carlo estimation \cite{numCsse}, though using a different method with free boundaries and a size-dependent coupling tuned to exhibit scale invariant correlations. To our knowledge, our estimation is the first to be directly in bulk marginals, and brings one additional order of magnitude in precision to the numerical check.

Another interesting element of the CFT data is the three-point energy field coefficient $C_{\varep \varep \varep}$. This vanishes because $\varep$ is odd under the duality operation of the Ising model swapping the spin and disorder fields. This duality maps the high-temperature expansion of the partition function to the low-temperature expansion. The critical point $\beta_c$ is self-dual, thus giving rise to an extra $\Z_2$ symmetry that is not apparent in the Ising measure or the  MCMC sampler. This duality is broken by the initial state of the Markov chain, and thus needs to be built by its mixing. We also note that the variable $\varep_{\bf i}$ transforms multiplicatively under duality only after appropriate shifting by the non-universal expectation value ${\cal H}$ \eqref{defener}, again pointing to the nontriviality of this extra $\Z_2$ symmetry. The check consists of a fit of the lattice energy density three point function on:
$$ \big\bra\  \varep_{\bf i }\,\varep_{\bf j }\,\varep_{\bf k} \ \big\ket \sim \frac{C_{\varep \varep \varep} }{|x-y| |y-z| |z-x|}\ a^3\,N_{\varep}^3$$
(again with $x=a{\bf i}$, $y=a{\bf j}$ and $z=a{\bf k}$), and using insertions with distances in a range from 7 to 40. This gives us the following numerical estimation:
$$ C_{\varep \varep \varep} = -0.0105\,(2).$$
Fitting a vanishing coefficient with more precision is a real challenge as it is dominated by noise. Here, our best effort gives us an estimation quite supportive of the duality invariance constraint.

\subsection{The holomorphic stress-energy tensor and the conformal Ward identities}

One of the most fundamental fields in CFT is the holomorphic stress-energy tensor (see e.g. \cite{df_cft}). Its existence indicates the presence of {\em local} conformal invariance: the fact that Ising measures defined on conformally equivalent domains (not necessarily simply connected), when conformally transported, give rise to the same scaling limits  (see \cite{ginsparg,df_cft} for a field-theoretic explanation of this relation, and \cite{sheffield,werner,doyon_random} for a conformal-loop-ensemble demonstration).

The main property of the stress-energy tensor is that it has dimension and spin $\Delta_T = s_T = 2$. It is in fact the leading coefficient with spin 2 in short-distance expansions of generic observables in CFT. For instance, the spin-spin operator product expansion, in the identity operator channel, gives \cite{df_cft}:
$$\sigma(x) \sigma(0) = \frac{1}{|x|^{\frac{1}{4}}}\big({\mathrm{I} + x^2 \frac{\Delta_{\sigma}}{c}T(0) + \mathcal{O}(x^3)\big)},$$
which can be inverted by means of a Fourier transform:
\beq\label{TCFT}
T(0) = \frac{c}{\Delta_{\sigma}} \;\lim_{r\to 0}\, \frac{1}{2 \pi} \int \mathrm{d}\theta \;\mathrm{e}^{-2i\theta}\,r^{-\frac{7}{4}} \;\sigma(r\mathrm{e}^{i \theta})\, \sigma(0).
\eeq
One can also argue that the holomorphic stress-energy tensor should be expected to be reproduced by the most local spin-2 lattice variable.

In order to define the stress-energy tensor on a triangular lattice, we may thus extend the definition used in \cite{numC} from square to triangular lattices:
\beq\label{Tlattice}
	T_{\bf i} = \sum_{{\bf j}\, \in \,{\cal N}({\bf i})} \text{e}^{-2i\theta_{\bf ij}}\sigma_{\bf i} \sigma_{\bf j}.
\eeq
Note that here (and contrary to the square lattice case) the variable $T_{\bf i}$ takes complex values.

Up to a normalization, this definition generalizes \eqref{TCFT} to the lattice, as the integral of $\theta$ becomes a sum over the symmetry directions. Since, as we saw, the spin-spin correlator exhibits a CFT behaviour already from a distance of 1, it indeed appears to be sufficient to take a sum over first neighbours in order to implement \eqref{TCFT} on the lattice.

A more refined definition might be obtained by using instead the next-to-nearest, or farther, neighbours. The price would be to losing locality, and thus increasing microscopic effects and requiring larger distances in correlation functions in order to reproduce CFT behaviours. However, clearly definition \eqref{Tlattice} has spin 2 under the lattice rotation symmetry group, and it is the most local spin-2 variable that can be constructed. Independently from its relation with the CFT formula, this definition may therefore be expected to reproduce the holomorphic stress-energy tensor. As we see below, this is indeed the case.

The relation with the CFT holomorphic stress-energy tensor $T(z)$ is then expected to be
\beq\label{Tiz}
	T_{\bf i} \sim N_T a^{2}T(z)\quad (a\to0,\;z=a{\bf i}\ \mbox{fixed}).
\eeq

We repeated the estimation for the power law $\Delta_T$ of the absolute value of the two-point correlator $|\bra T_{\bf i} T_{\bf j}\ket|$. This numerical estimation was significantly more challenging than in the cases of the spin or energy variables, since its correlations decay much faster with distance (as $|z|^{-4}$), the signal being rapidly dominated by noise. We had to extend to approximately 670 000 uncorrelated lattice samples. On these, we looked in a restricted subdomain excluding 30 lattice sites from the boundaries, and used operator insertions along three different directions $\theta_{{\bf ij}} \in \{\pm\frac{\pi}{3},0\}$. This choice was motivated by computational efficiency. See Fig. \ref{figTT}.
\bfgr
\begin{center}
\includegraphics[width=.8\lw]{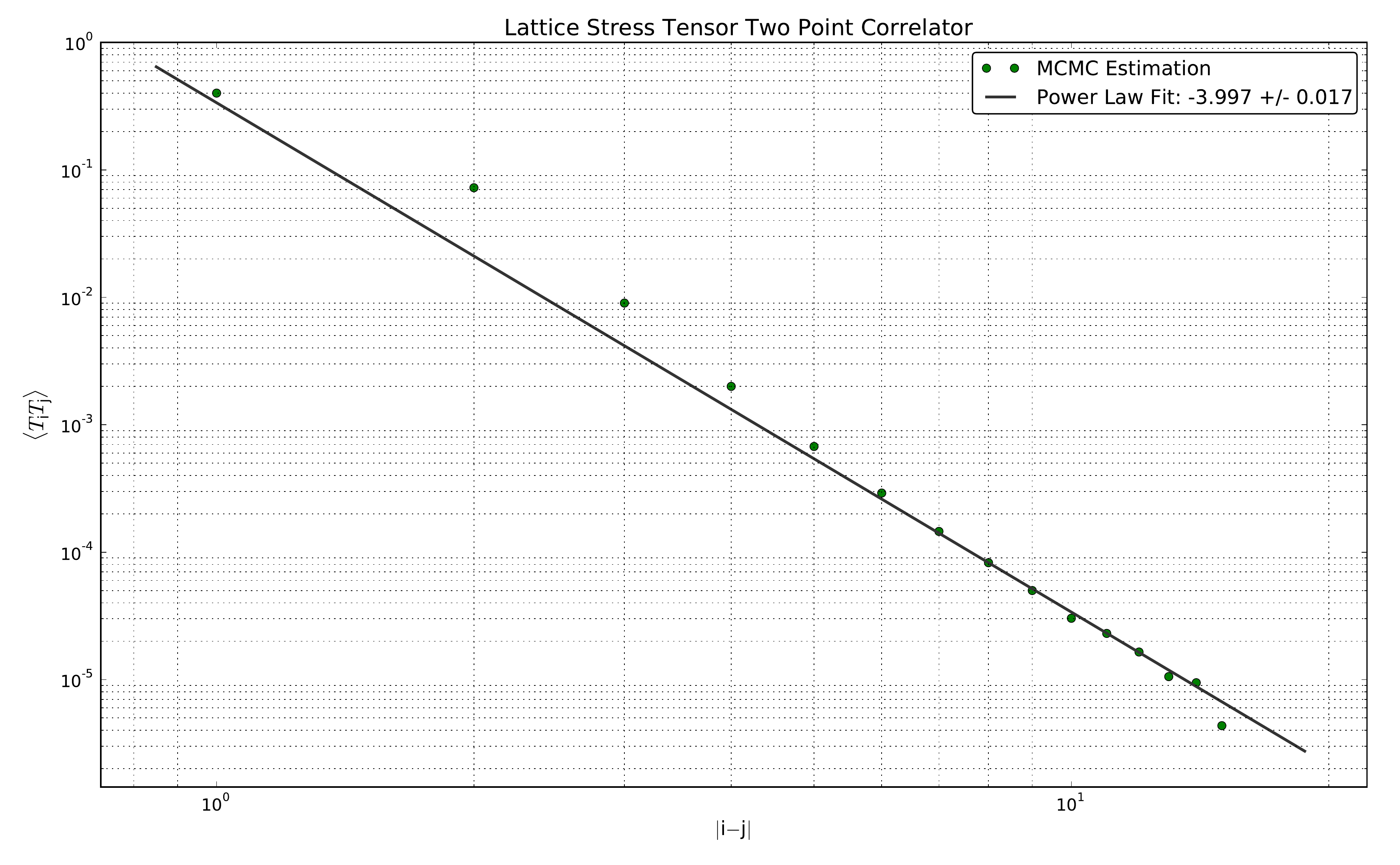}
\captionsetup{singlelinecheck=off}
\caption[foo bar]{\hspace{.4cm}\small{Absolute value of the lattice stress-energy tensor two-point correlator. We see three different ranges:
\begin{itemize}
\item At short distances - separations less than 6 - the signal seems to suffer from an excess of correlations. We attribute this to microscopic effects. This range of microscopic effects is similar to that found in the energy correlator $\bra \varep_{\bf i}\varep_{\bf j}\ket$.
\item We see a good power law profile up to distances $\approx 15$, where we fitted the exponent and the offset using the subset of points minimizing the $\chi^2$ of the fit of the power law.
\item Beyond that, the signal has already decayed and becomes dominated by the noise.
\end{itemize}
}}
\label{figTT}
\end{center}
\efgr

We clearly see a `bump' above the fitted power law behaviour, very strong at distance 2 (note that ${ T}_i$ and ${T}_{i+2}$ overlap by sharing a lattice spin) and disappearing for distances larger than 7. Because of the difficulty we had in fitting this data on a power law, we decided to add a prescription in the fit and only include the set of points inducing the smallest uncertainty on the fitted exponent value, or in other words the distances for which the power law behaviour was the most manifest. This gave us points at distances 8, 11 and 12; for a fitted exponent value of $\Delta_{{\cal T}} = 1.998 \pm 0.009$. This is in satisfying agreement to the CFT value.

We also repeated the estimation of the spin $\sigma_T$. Instead of looking at directions $\{ \pm\frac{\pi}{3}, 0\}$, we decided to perform this consistency check by looking at its correlation to its neighbours in the directions $\{\pm\frac{\pi}{6}, \frac{\pi}{2}\}$  - including neighbours with distance up to 11. Splitting the real and imaginary part and fitting them on a cosinus and sinus respectively gave us the numerical spin value of $2.0011 \,(5)$, in good agreement with the CFT expectation.

\subsection{``Numerical'' central charge}

Although not identifying it uniquely, the central charge is a key characteristic of a CFT model: for instance, its monotonic behaviour under RG flows has clear physical interpretations, and its involvement in the Virasoro Algebra encodes how descendant operators transform under conformal mappings. However, numerically, it is harder to unravel as it is not a direct observable.

Different methods for computing it have been used in the past: for instance, putting the theory on a cylinder and looking at the scaling behaviour of the free energy, or, more recently, using the twist operator in order to construct conical singularities (related to the entanglement R\'enyi entropy).

Here, we derive it in bulk marginals simply by continuing our study of correlation functions. It is known from CFT that the stress-energy tensor has the following two point correlator:
 $$\langle\; T(x)\,T(y)\;\rangle=\frac{c}{2}\,\frac{1}{(x-y)^4}.$$
In order to use this formula, we must numerically estimate the normalization constant $N_T$ in \eqref{Tiz}. We therefore first look at the stress-energy tensor insertion within a two-spin correlator, which by Ward identities must obey:
\begin{equation*}
\begin{split}
\langle\; T(x) \,\sigma(y)\, \sigma (z) \;\rangle 
 & = \frac{h_{\sigma}}{|y-z|^{\frac{1}{4}}} \frac{(y-z)^2}{(x-y)^2(x-z)^2}
\end{split}
\end{equation*}
where $h_{\sigma} = \frac{1}{16}$ is the scaling weight of the holomorphic part of the spin operator.
 
In order to evaluate $\langle{ T}_{\bf i}\,\sigma_{\bf j}\,\sigma_{\bf k}\rangle$, we assumed invariance under translations and restricted ourselves to horizontal insertions (for simplicity, and thus only keeping a dependence on two distances). The lattice stress-energy tensor is inserted in-between the spins with a maximal separation of 45 lattice units to each of them. The insertions were also constrained to a minimal distance to the borders of 40 so as to avoid boundary induced departures from the bulk marginal. The estimation ran on 3 000 samples of size 512x512. See Fig. \ref{3pttssv2}.
\bfgr
\begin{center}
\includegraphics[width=.9\linewidth]{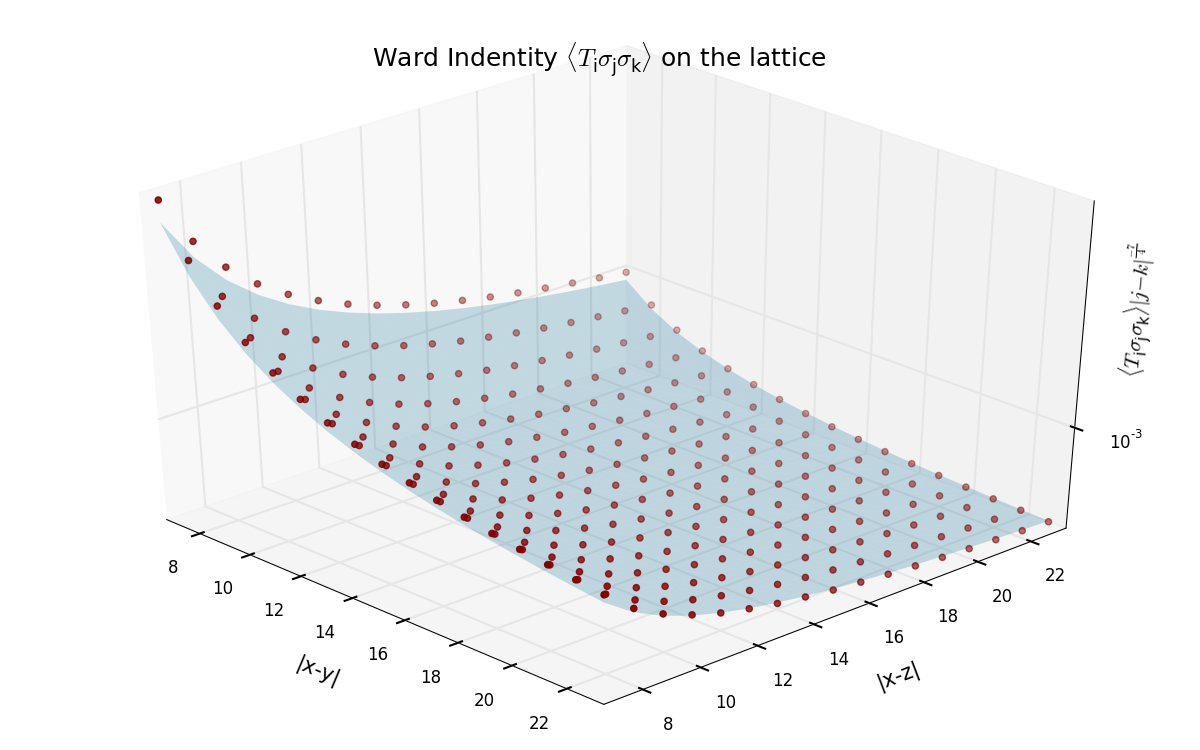}
\caption{\small{This 3d plot of $\bra T_{\bf i} \si_{\bf j} \si_{\bf k} \ket \,|{\bf j}-{\bf k}|^{-\frac{7}{4}}$ versus $|{\bf i} - {\bf j}|$ and $|{\bf i} - {\bf k}|$ offers a graphical estimation of the superposition to $a^4\,N_{\cal T} N_{\si}^2 \,\bra T(x) \si(y) \si(z) \ket\, |y-z|^{-\frac{7}{4}}$ for $N_{T} N_{\si}^2 = 1.29$. Points are kept on a horizontal line, avoiding angular dependences. The agreement of our collected data points to the theoretical expectation appears to be quite good, especially in the region of separations $|{\bf i} - {\bf j}|,|{\bf i} - {\bf k}| \geq 11$.}}
\label{3pttssv2}
\end{center}
\efgr

The proportionality  between the two graphs is apparent (as well as the overall lack of noise in retrieving this pure CFT result). We decided to fit the coefficient $N_{\si}^2 N_{T}$ by optimization, looking at the minimum of:
$$\sum_{{\bf j},{\bf k}} \bigg(\big\langle{ T}_{\bf{i}}\,\sigma_{\bf{j}}\sigma_{\bf{k}}\big\rangle  - a^{\frc94}\,N_{\si}^2 N_{T} \;\big\langle T(x)\,\sigma(y)\sigma(z)\big\rangle\bigg)^2$$
with, again, $x=a{\bf i}$, $y=a{\bf j}$ and $z=a{\bf k}$. The sum is restricted to points with $\big\langle{ T}_{\bf{i}}\,\sigma_{\bf j}\sigma_{\bf k}\big\rangle > 0.0015$, thus excluding points more sensitive to noise, and also leaving out distances between insertions closer than 3 lattice units, as they are too sensitive to the microscopy. The above quantity has a paraboloid aspect (see Fig. \ref{paraboleNt}) and shows a clear minimum at
\beq\label{NsiNT}
	N_{\si}^2 N_{T} = 1.29\,(1).
\eeq
Using \eqref{Nsi}, this gives $N_T = 1.93\,(1)$.

\bfgr
\begin{center}
\includegraphics[width=.8\linewidth]{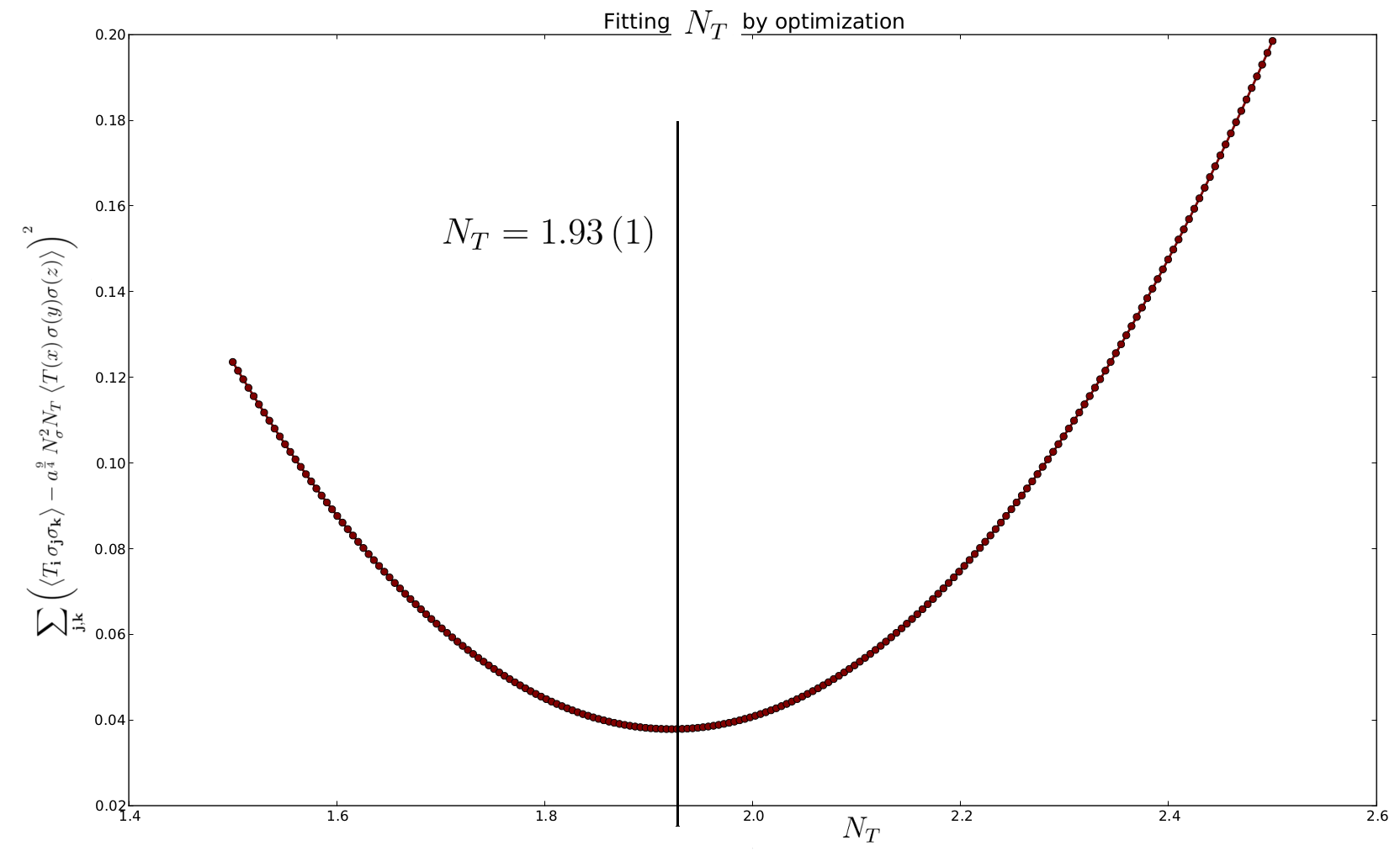}
\caption{\small{Plotting $f(N_T) = \sum_{{\bf j},{\bf k}} \Big(\big\langle{ T}_{\bf{i}}\,\sigma_{\bf{j}}\sigma_{\bf{k}}\big\rangle  - a^{\frc94}N_T N_\sigma^2\;\big\langle T(x)\,\sigma(y)\sigma(z)\big\rangle\Big)^2$. The clear parabola profile offers a global minima at $N_T N_{\si}^2 = 1.29(1)$ using BFGS numerical minimization method. Using our previous numerical estimation of $N_{\si}$, we derive $N_T = 1.93\,(1)$.}}
\label{paraboleNt}
\end{center}
\efgr
We may now evaluate the central charge by estimating the amplitude of the power law $\langle{ T}_{\bf i}\,{ T}_{\bf j}\rangle \sim \frc c2 N_T^2 ({\bf i}-{\bf j})^{-4}$. In order to obtain the most precise fit of the corresponding offset in a log-log fit, we decided to run the calculation of the correlator from scratch on a set of approximatively 500 000 samples, excluding as far as 50 lattice units from the boundaries and looking at distances between insertions of 5, 6, 7 and 8.

The fit of the offset gives us
\beq
	\frac c2 N_{\cal T}^2 = 0.920 \pm 0.006
\eeq
(fitting between distances 4.72 and 8, for which the goodness of the fit is optimal). Using \eqref{Nsi} and \eqref{NsiNT}, we get the following numerical estimation for the central charge:$$c_{\text{numerical}}=0.493\pm 0.008.$$ This estimate is not directly centered on the theoretically expected value $c_{\mathrm{Ising}} = 0.5$, but is within one standard deviation. To our knowledge, at least one comparable numerical study shows a slightly more precise estimation of the quantity \cite{numC}, by using a finite size scaling method on the torus.
 
\subsection{Four-point functions}

The quantities that encode most of the information of a CFT model are the four-point correlation functions. The lattice spin four-point function is known from CFT\cite{ginsparg} to have the following analytical form, solution of the corresponding second order differential `null state' equation specific to the minimal model describing the critical Ising model:
$$\big\bra \si(z_1)\si(z_2)\si(z_3)\si(z_4)\big\ket = \frac{1}{\big|z_{12}z_{13}z_{14}z_{23}z_{24}z_{34}\big|^{\frac{1}{12}}}\, F_{\si\si\si\si}(\eta, \bar{\eta})$$
where $z_{ij} = z_i - z_j$, $\eta = \frac{z_{12}z_{34}}{z_{13}z_{24}}$, and the ``dynamical factor'' is:
$$F_{\si \si \si \si} (\eta, \bar{\eta}) = \frac12 \,\big|\eta (1-\eta)\big|^{-\frac16} \,\big( \big| 1- \sqrt{1-\eta} \big| +  \big|1+ \sqrt{1-\eta}\big|\big).$$

This ``signal'' does not decrease as quickly as the energy or stress-energy correlators but the challenge appearing here is the large freedom in which we can make our insertions. Our first numerical estimation was carried in the limit where two pairs of close insertions are well separated, here $|z_{12}|,|z_{34}|\ll|z_{14}|,|z_{23}|,\ldots$. The four point function simplifies to:
$$\big\langle \sigma(z_1)\,\sigma(z_2)\,\sigma(z_3)\,\sigma(z_4)\big\rangle \; \approx \; \frac{1}{|z_{12}|^{\Delta_1 = \frac{1}{4}}|z_{34}|^{\Delta_2 = \frac{1}{4}}}.$$

On a sample of 512x512 sublattices we estimated the quantity using insertions at least 60 lattice units from the boundaries and pairs within which the separation did not exceed 60 lattice units and respectively separated by at least 160 lattice sites. We also restricted ourselves to horizontal directions for simplicity. Fitting our data on a double power law gave us the following results for different cut off distances inside the two pairs:
\begin{center}
\begin{tabular}{c|c c}
Cut-Off & $\Delta_1$ & $\Delta_2$ \\ \hline \\
60 & $0.2525 \, (3)$ & $0.2548 \, (4)$ \\
40 & $0.2517 \, (2)$ & $0.2532 \, (4)$ \\
20 & $0.2503 \, (2)$ & $0.2513 \, (2)$ \\
5 & $0.2502\, (4)$ & $0.2503 \, (4)$ \\
\end{tabular}
\end{center}

We clearly see that as we reduce the accounted contributions to the closest insertions within the pairs, the fit becomes more and more precise.\\

Beyond this ``dipolar'' approximation, we ran two estimations of $F_{\si\si\si\si}$ using different prescriptions in inserting the four spins operators:
\begin{itemize}
\item First we used 100 0000 random insertions inside a sample of $\sim$ 500 000 sublattices of size 512x512. We ran a $\chi^2$ estimation summing over the set of insertions to find $\frac{\chi^2}{\mathrm{d.o.f.}} \approx 5.10^{-3}$ and a p-test value of 1. Restricting to insertions no more than 100 lattice units apart the normalized $\chi^2$ is evaluated to $6.10^{-4}$. This number supports more firmly of a good agreement of our sample to the theory.
\item Second we used insertions along the same -- horizontal -- direction to measure the cross ratio $\eta$ on the real line but over an interesting range: $\eta \rightarrow 0,1, +\infty$ (the last one is obviously unreachable numerically). This gave us Fig. \ref{4ptssss}.
\end{itemize}
\bfgr
\begin{center}
\includegraphics[width=.9\linewidth]{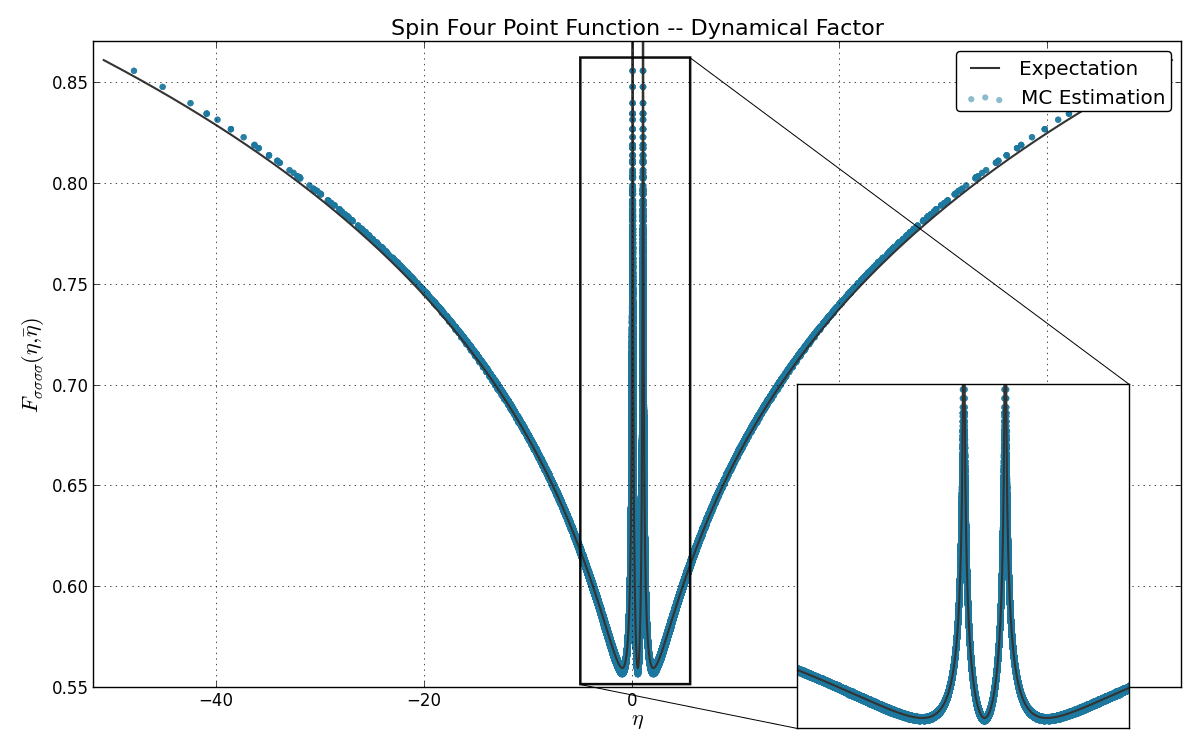}

\caption{\small{Plotting the dynamical factor $F_{\si\si\si\si}(\eta,\bar{\eta})$ versus $\eta$ with the CFT result (grey line) and our numerical estimation (blue points). In this range of $\eta$, real as the positions are taken to be horizontally aligned, we cover most of the interesting behaviour of $F_{\si\si\si\si}$. The agreement is manifest, with a normalized $\chi^2$ of 0.0354 and a p-test value of 1. The most interesting behaviour of this function happens for $\eta \to 0,1,\pm \infty$. The zoom on the graph shows the great agreement of our measurements at the former two points. Our large-$|\eta|$ values also seem to exhibit the right asymptotic behaviour.}}
\label{4ptssss}
\end{center}
\efgr

To our knowledge these are the first numerical checks of a CFT four-point functions in the planar critical Ising model.

We extended this work to the four-point functions $\bra \varep \varep \si \si \ket$ and $\bra \varep \varep \varep \varep \ket$ given by \cite{corr_Mattis}:
\begin{eqnarray*}
\bra \varep(z_1) \varep(z_2) \si(z_3) \si(z_4) \ket &=& \frac{|z_{34}|^{\frac{7}{4}}}{|z_{14}|^2  |z_{23}|^2}\; F_{\varep\varep\si\si}(\eta,\bar{\eta}),\\
F_{\varep\varep\si\si}(\eta,\bar{\eta}) &=& \frac{1}{4} \Bigg| \frac{(\eta-2)\sqrt{\eta-1}}{\eta}\Bigg|^2
\end{eqnarray*}
and
\begin{eqnarray*}
\big\bra \varep(z_1) \varep(z_2) \varep(z_3) \varep(z_4) \big\ket &=& \bigg|\frac{1}{z_{14}z_{23}}\bigg|^2\;F_{\varep\varep \varep\varep}(\eta,\bar{\eta}),\\
F_{\varep\varep\varep\varep}(\eta,\bar{\eta}) &=& \Bigg| \frac{\eta^2-\eta+1}{\eta} \Bigg|^2.
\end{eqnarray*}
We offer numerical checks in Fig. \ref{4pteess} and Fig. \ref{4pteeee} respectively.
\bfgr
\bc
\ig[width=.75\lw]{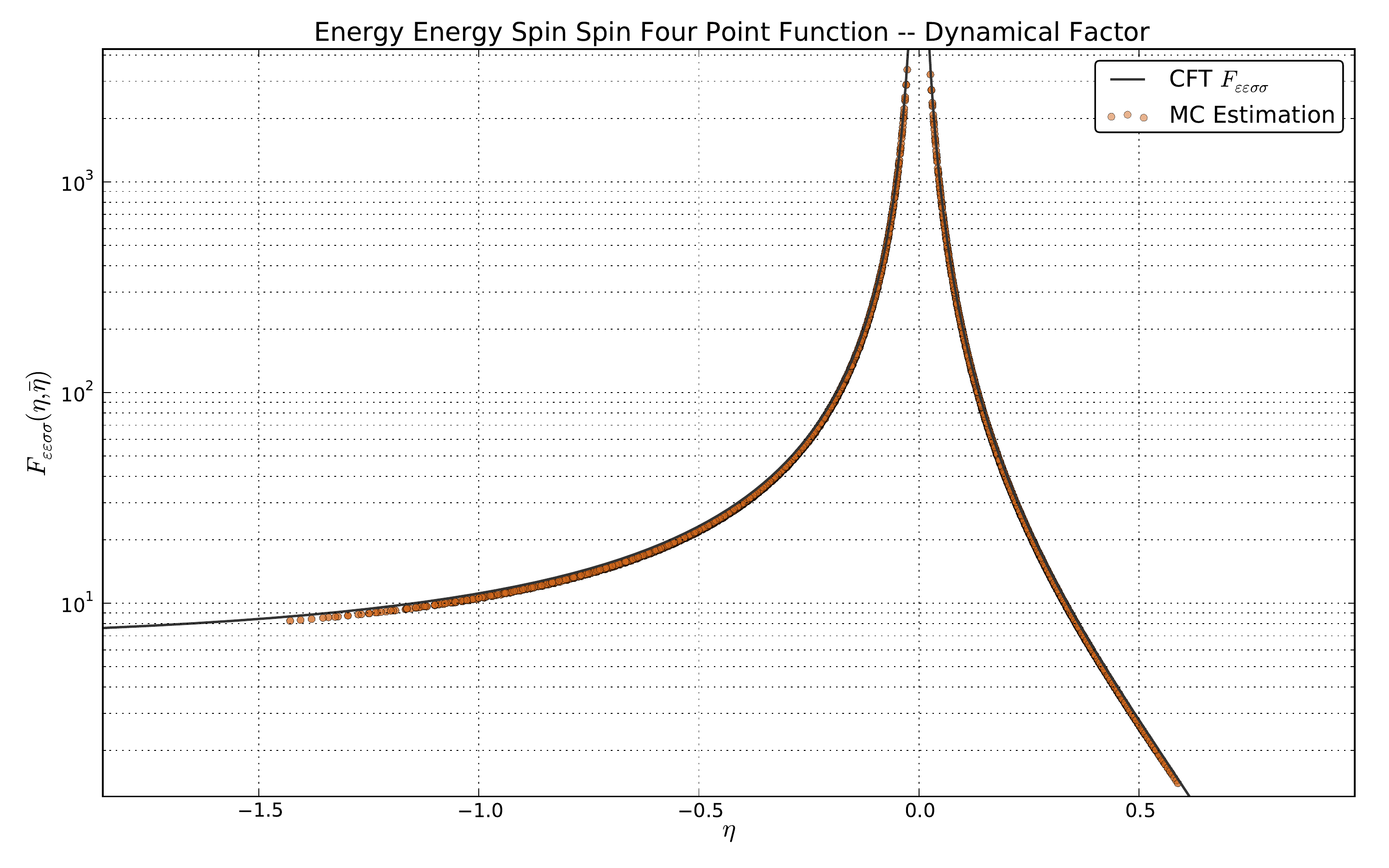}
\caption{\small{Dynamical Factor of the $\bra \varep \varep \si \si \ket$ four point function. The agreement is satisfying: the $\chi^2$ of our estimator is 1.59 with a p-value of 1.}}
\label{4pteess}
\ec
\efgr

\bfgr
\bc
\ig[width=.75\lw]{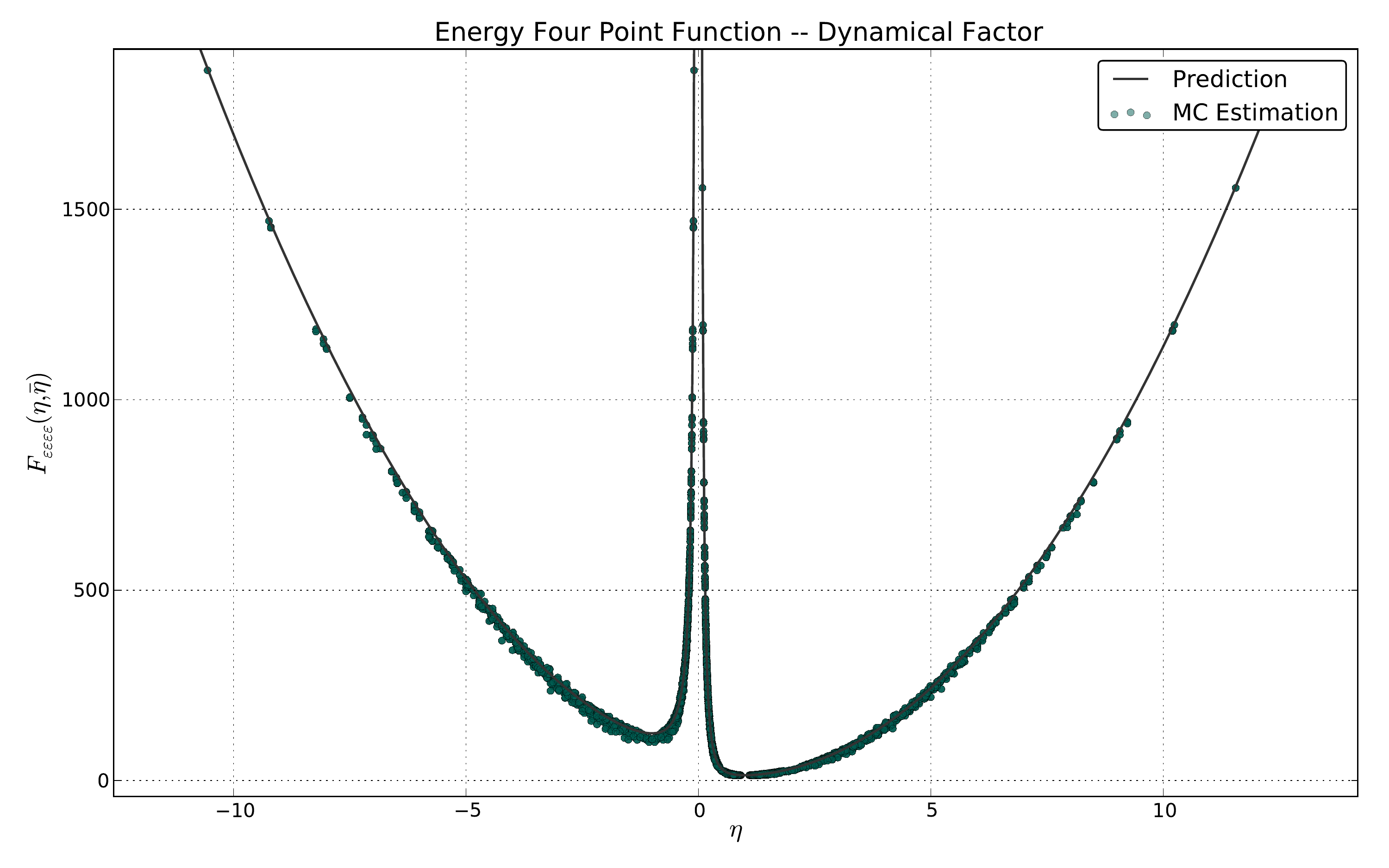}
\caption{\small{Dynamical Factor of the $\bra \varep \varep \varep \varep \ket$ four point function versus crossing ratio $\eta$, here only taking real values. We see an obvious agreement on this range of $\eta$ covering the most interesting behaviour of $F_{\varep\varep\varep\varep}$. The $\chi^2$ is here 164.14 and the p-test value of 1, supporting the obvious agreement between our estimations points and the prediction line.}}
\label{4pteeee}
\ec
\efgr

\section{Conclusions and outline}
\label{sectconclu}

In this work, we introduced a numerical method for sampling finite-domain marginals of infinite-volume critical systems (bulk marginals). The idea of the method is to send, using scale invariance, any initial boundary information to infinity. This is done by constructing, on the boundary of the finite domain of which the marginal is taken, a new ``holographic'' boundary condition that encodes the infinite volume beyond it. From the viewpoint of the renormalization group, the method provides a numerical way of approaching the UV fixed point, effectively zooming onto small regions.

The method was implemented as a Markov chain Monte Carlo sampler, which we refer to as a UV sampler, for the planar Ising model. We checked the quality of the resulting samples by verifying scale invariance of the spin-spin correlations, and we verified that there was satisfactory agreement with known analytical predictions of various numerical values of the underlying infinite-plane CFT data (including three- and four-point function, the central charge and the conformal Ward identities). We also perfomed checks on the global loops distribution \cite{loopDistriToAppear}.

The residual effects of the bulk marginal boundary were also analyzed. These were observed to give rise, to a good approximation, to power laws with small amplitudes. The powers observed depend continuously on the scaling parameter used to zoom in. This was tentatively interpreted in terms of the density of holes on the boundary, which necessarily appear in the numerical procedure due to the partial information about the UV fixed point. It would be interesting to have a full understanding of these residual power laws.

Although we presented results for bulk marginals on square domains, the UV sampler can be implemented for any shape of the domain boundary, and we also implemented it on circular domains. The small anisotropies induced by the square boundary are further reduced, and this might open the way to higher-precision fitting of bulk-marginal data. However, we have not yet fully solved the problem of establishing a uniform distribution of holes on the circular boundary.

In the context of two-dimensional CFT, numerical transfer matrix methods have been extensively and successfully used in order to study planar two- and three-point functions with high precision (see for instance \cite{transferMatrix1, transferMatrix2}). These methods are based on the relation that exists between distributions on the plane and on the cylinder, thanks to two-dimensional conformal invariance. They have the advantage, over Monte Carlo samplers, of retaining the full information of the distribution. On the other hand, transfer matrix methods are strongly limited in the size of the system, while the UV sampler can generate bulk marginals on much larger lattices. Hence, we expect the UV sampler may provide more precise numerical estimations,  unaffected by microscopy, for properties of extended objects such as random cluster boundaries. In addition, the UV sampler seems to offer much more flexibility, and in particular does not require, for its basic working principles, the strong conformal invariance of two-dimensional CFT.

Many generalizations of the present work are possible (in particular, the last two are not accessible with transfer matrix methods):
\bi
\item[$\ast$] The method is not limited to holographic boundary conditions representing an empty infinite volume beyond it: the general CFT arguments presented show that, with simple modifications, one may construct a holographic boundary condition representing the insertion of any number of local observables beyond it. This may be useful to study correlation functions and extract high-precision exponents.

\item[$\ast$] Generalizations to Potts models or more general lattice models with finite-range interactions are expected to be straightforward. The present proposal can be extended to critical $O(n)$ loop-gas models, where the Hamiltonian becomes severely nonlocal in terms of underlying spins. We have a paper in preparation where the UV sampler for the critical $O(n)$ model is studied, generating bulk marginals with a similar efficiency.

\item[$\ast$] It is possible to extend the UV sampler to the sampling of bulk marginals for massive QFT -- that is, in the near-critical scaling limit. Indeed, dilations are not requested to be fixed point, and one may simply adjust the lattice coupling after a dilation procedure. Recall that the correlation length $\xi$ scales as $\xi \sim |T-T_c|^{-\nu}$ with the temperature, and recall that a dilation by $\lambda$ should send $\xi \rightarrow \lambda \xi$. Then, the initial configuration is a sample with $T$ away from $T_c$, and we simply do the replacement $|T-T_c|\rightarrow |T-T_c| \lambda^{-\frac{1}{\nu}}$ after every blow-up step, thus approaching the critical temperature $T_c$. We stop when the resulting $\xi$ is of the order of the lattice size, thus obtaining a bulk marginal with finite correlation length.

\item[$\ast$] Finally, perhaps the most interesting generalization is to higher dimensions. The three-dimensional critical Ising model is a popular toy model for studying three-dimensional CFTs, and many numerical achievements have been made over the last few years \cite{ising3d1,ising3d2}. Generalizing the UV sampler to the 3d Ising model is straightforward and we have preliminary results showing that it works with comparable efficiency. A paper is being prepared.

\ei

\subsubsection*{Acknowledgements}

We would like to thank G. M. T. Watts and Y. Ikhlef for fruitful discussions during this research. BD thanks EPSRC for support at the beginning of this project under the grant ``From conformal loop ensembles to conformal field theory'' EP/H051619/1. VH is funded by a Graduate Teaching Assistantship, KCL Department of Mathematics.

\newpage

\appendix
\section{Swendsen-Wang algorithm with fixed boundaries}
\label{sec:appendixSW}

Critical slow down in numerical simulations of statistical systems is a direct consequence of the scaling invariance of the system and of the inexistence of a lower bound in the spectrum of the excitations. Close enough to the critical point, the correlation length becomes much larger than $L$ the linear size of the lattice put on a computer. This means that the Monte-Carlo evolution of the system will have to update very low energy modes, modes of order $\sim \frac{1}{L}$.

With local updates (such as single spin Metropolis, Glauber, heat bath, ...),  the chain autocorrelation time $\tau_{{\rm ac}}$ is known to grow as $L^{4}$. This makes precise Monte-Carlo estimations of global observables much difficult; recalling that the uncertainty of an observable $\Or$ goes as:
$$ {\delta \Or}^2 \approx \frac{\tau_{ac}\times {\rm variance}(\Or)}{\text{computation time}}.$$

Since the seminal paper of Swendsen and Wang \cite{swendsenWang} the way to circumvent this obstacle is to apply non local updates, updating large fractions of the lattice at once. On the Ising model, the implementation came from the use of the Fortuin-Kasteleyn (FK) formulation of the partition function \cite{FKclusters1,FKclusters2} in terms of clusters of spins sharing the same sign, bonded one to each other with a probability:$$p_{\rm bond} = 1 - e^{-2\beta_c}.$$

From the partition function of the Ising model defined on spin configurations:
\begin{eqnarray}
Z_{\rm Ising}(T=T_c, h = 0) &=& \sum \limits_{\{ \si \}} e^{\beta_c\sum_{({\bf i}, {\bf j} \in {\cal E})} \si_{\bf i } \si_{\bf j} - 1}\nonumber \\
&=& \sum \limits_{\{\si\}} \prod \limits_{({\bf i}, {\bf j}) \in {\cal E}} \Big((1-p) + p \delta_{\si_{\bf i }, \si_{\bf j}}\Big) \qquad {\rm with} \quad p = p_{\rm bond}\label{ZisingToFK}\\ 
&=& \sum \limits_{\{\si\}} \sum \limits_{\{n_{\bf ij}=\, 0,1\}} \prod \limits_{({\bf i },{\bf j}) \in {\cal E}} \Big((1-p)\delta_{n_{\bf ij},0} + p \delta_{\si_{\bf i},\si_{\bf j}} \delta_{n_{\bf ij, 1}}\Big).\label{ZisingBondsSpins}
\end{eqnarray}
In the last line we used the trivial identity $x + y = \sum\limits_{n=0}^1\; x \delta_{n,0} + y \delta_{n,1}$.

Relation (\ref{ZisingBondsSpins}) gives a partition function written in terms of two sets of local variables: spins $\{ \si\}$ and bonds $\{ n \}$. From the partition function we read a dictionary between the two. Given a configuration of spins $\{ \si \}$, we have the following distribution for $n_{\bf ij}$:
\begin{align*}
{\rm if}\ \si_{\bf i } = \si_{\bf j} &:&p(n_{\bf ij} = 1) = p \quad {\rm and} \quad  p(n_{\bf ij} = 0) = 1- p(n_{\bf ij} = 1) = 1-p,\\
\textrm{else} &:&p(n_{\bf ij} = 1) = 0 \quad {\rm and} \quad p(n_{\bf ij} = 0) = 1.
\end{align*}
It reads that a bond $n_{\bf ij}$ can be ``activated'' with probability $p$ only if it links two neighbour spins of identical signs. This allows to write a configuration of bonds from a configuration of spins.

In the other direction, given a configuration of bonds $\{n\}$, we have for two neighbours sites $\si_{\bf i}$ and $\si_{\bf j}$:
\begin{align*}{\rm if}\ n_{\bf ij} = 1&:& p(\si_{\bf i} = \si_{\bf j}) = 1\\
\textrm{else if}\ n_{\bf ij} = 0&:& p(\si_{\bf i} = \si_{\bf j}) = \frac12.
\end{align*}

This means that ``bonded'' spins ($n_{\bf ij} = 1$) must share the same sign while two ``unbonded'' spins ($n_{\bf ij} = 0$) have only a half probability to share the same value.

The duplicate description provided by (\ref{ZisingBondsSpins}) is employed in the Swendsen-Wang (SW) lattice update algorithm. An evolution step consists in covering the lattice with ``virtual FK clusters'' made of bonds joining neighbour spins of identical value and activated with probability $p$. Within each of these clusters, all spins are assigned the same value picked at random in $\{1,-1\}$ with equal probability. By going back and forth between the spin and the bond representation the method samples the Gibbs measure of the Ising model.

At the critical point it is of primary interest since it has been shown to counter significantly the critical slow down. The autocorrelations of this Markov Chain have been measured to scale as \cite{newman}:
$$\tau_{\rm ac} \propto L^{\approx 0.4}.$$

In the case of a lattice subsection $A$ with a fixed boundary $\partial A$, here fixed can mean entering the Gibbs measure but non dynamical with respect to the Markov Chain. Because of its locality, the Hamiltonian decomposes as:
$$H_{\cl A} = H_{\p A} + H_{I(A)},$$
where the first term is the energy stored on the edges connecting the boundary spins $\si_{\p A}$ to the subsection inside A:
$$H_{\p A}(\si_A | \si_{\p A}) = -\sum \limits_{( {\bf i} \in \p A, {\bf j} \in A )} \si_{\bf i} \si_{\bf j}-1,$$
while the second term is the energy contribution restricted to the inside of the lattice $I(A) = \bar{A} \setminus \p A$:
$$H_{I(A)}  = -\sum \limits_{( {\bf i},{\bf j} ) \in {\cal E_A}} \si_{\bf i} \si_{\bf j}-1.$$

Introducing the boundary contribution of the energy in \eqref{ZisingToFK}, the bottom line reads:

$$Z_{\rm Ising}(\beta_c,\si_{\p A}) = \sum \limits_{\{n\}} \sum \limits_{\{\si\}} \bigg( \prod \limits_{({\bf i},{ \bf j}) \in {\cal E}_A}(1-p)\delta_{n_{\bf ij},0} + p \delta_{\si_{\bf i}, \si_{\bf j}} \delta_{n_{\bf ij},1} \bigg) e^{-\beta_c H_{\partial A}({\si},\si_{\p A})}$$

In the implementation of SW lattice flips with an external magnetic field $h$ \cite{wolff_bnd}, the sign attributed to a virtual cluster $\cal C$ is no longer equiprobably $\pm 1$ but must obey a heat-bath assignation:
$$p(\si_{\cal C} = \pm) \propto e^{\pm\beta h |{\cal C}|}$$
where $|{\cal C}|$ is the mass of $\cal C$, e.g. how many spins it contains.

In the same spirit, after having covered $A$ with virtual FK clusters $\{ {\cal C}_i\}$, looking individually at each cluster ${\cal C}_i$:

\bi
\item if none of its spins $\si_{\bf k}$, ${\bf k} \in {\cal C}_i$, share a bond with the boundary spins $\si_{\p A}$ then it can be assigned a random value $\pm 1$ with identical probabilities.
\item otherwise, if some of its spins share a bond with $\si_{\p A}$, an energy contribution on the boundary can be calculated:
$$H^i_{\p A} ( \pm , \si_{\p A}) = - \sum \limits_{({\bf i} \in \partial A, {\bf j} \in {\cal C}_i)} (\pm) \si_{\bf j} - 1$$
summing the energy stored in the bonds between ${\cal C}_i$ and $\p A$. This contribution can now be used in a heat bath assignation method for ${\cal C}_i$:
$$p(\si_{{\cal C}_i} = \pm) \propto e^{\beta_c H^i_{\p A} (\pm, \si_{\p A})}$$
\textit{Equivalently}, for computational effectiveness this could be replaced by Metropolis-like acceptance ratio where each ${\cal C}_i$ is flipped ($\si_{\bf j} \rightarrow - \si_{\bf j}, \forall {\bf j} \in {\cal C}_i$) with probability:
$$p_{\rm acc} = {\rm min}\Big(\,1,\, e^{\beta_c \big(H^i_{\p A}(\si_{{\cal C}_i}, \si_{\p A}) - H^i_{\p A}(-\si_{{\cal C}_i}, \si_{\p A})\big)}\Big)$$
\ei

%
%
%
%

Wolff cluster flips could also be used in the same fashion with fixed boundaries but our experience is that the rejection rate will be very high for the clusters nearing the borders and these spins will be much harder to update. Wolff updates could also be coupled to Metropolis flips localised on a border crown, because of the strong critical slow down of single flips, this would not be as efficient as using Swendsen-Wang algorithm. On top of that, the scaling behaviour of SW with the lattice size is well known. In our chain we checked that the number of lattice flips to achieve thermalization is constant with the lattice size or the dilation parameter. Though, autocorrelations of the magnetization and the energy do seem to scale as $L^{0.4}$, with $L=h=v$ e.g. picking square subsections.

\section{A CFT analysis of the lattice Markov chain}\label{appCFTmarkov}

In this subsection we provide an elementary CFT analysis of the small-scale corrections in \eqref{latsca2}.

The large-distance asymptotic of correlation functions on a critical lattice are described by CFT. These leading asymptotic behaviors receive microscopic corrections due to the lattice, akin to those in \eqref{latsca2}. It is possible to describe such corrections within CFT, by considering a modification of the CFT action by irrelevant operators.

The principle can be illustrated for a two-point correlation function. Let $s_{\bf i}$ be the lattice variable corresponding to the CFT field $\Or(x = a {\bf i})$. Then the leading large-distance behavior of the lattice correlation functions of $s_{a \bf i}$ is reproduced by putting the fields $\Or(x)$ at the positions $x=ai$ where the lattice variables lies ($a$ is the lattice spacing), and evaluating appropriately normalized CFT correlation functions. This principle extends to the subleading behaviors by adding homogeneous terms to the CFT action which are integration of irrelevant operators, and evaluating the correlation function perturbed by such terms. For instance, with only one perturbation term,
\beq
	\bra s_{\bf i} s_{\bf j}\ket = a^{2\Delta_\Or}\bra \Or(a{\bf i})\Or(a{\bf j})\ket_g
\eeq
where $\Delta_\Or$ is the scaling dimension of $\Or$, and where $\bra\cdot\ket_g$ is the perturbed correlation function evaluated with the action
\beq
	S_g = S + g\int d^2x\,\phi(x) + \cdots
\eeq
with $\phi$ an irrelevant (in the RG sense) perturbing field, of scaling dimension $\Delta_\phi>2$. These expressions are formal, and must be interpreted within the rules of conformal perturbation theory. For instance, the first correction is obtained by expanding in $g$ to first order:
\beq\label{tp}
	\bra s_{\bf i}s_{\bf j}\ket = |{\bf i-j}|^{-2\Delta_\Or}\lt(
	1+ a^{2-\Delta_\phi}g \,C_{\Or\Or\phi}\,|{\bf i-j}|^{2-\Delta_\phi} {\cal I}
	+\ldots
	\rt)
\eeq
where $C_{\Or\Or\phi}$ is the three-point CFT coupling and ${\cal I}$ is evaluated by analytically continuing in $\Delta_\phi$ the integral
\beq
	\int d^2x\,\frc1{(|1-x|\,|x|)^{\Delta_\phi}}.
\eeq

Let us consider the Markov chain described in the previous sections, as it acts on the action describing the model. We start on a region $C$ with conformal boundary conditions, with CFT action $S_C$ and lattice-corrected action
\beq
	S_{C,g}=S_C + g \int_A d^2x \,\phi(x).
\eeq

For the sake of the discussion we invert the steps in one iteration, doing first the blow-up and then the re-thermalization, but the result is equivalent.

First we perform a blow-up step. The resulting measure on the lattice of {\em assigned spins}, with larger effective lattice spacing $a\mapsto \lambda a$ and larger domain $C\mapsto \lambda C$, has the same correlation functions, as functions of the new lattice positions, as those of the original measure on the spins in $A$. Therefore, from \eqref{tp} for instance, we see that the CFT action describing it is $S_{\lambda C, g\lambda^{\Delta_\phi-2}}$. This new CFT action is of course not constrained to the lattice of assigned spins, and thus immediately gives not only the measure on the lattice of assigned spins, but also that on all lattice spins in the larger region $\lambda C$. That is, it automatically encodes a certain prescription for filling-in the holes, which, we expect, is theoretically the most accurate.

Second we re-thermalize the subdomain $C$. Here we assume that re-thermalization is perfect in $C$, and thus the perturbation term recovers its original form in $C$. Hence we have
\beq
	S'=S_{\lambda C} + g\int_C d^2x\,\phi(x) + g\lambda^{\Delta_\phi-2}
	\int_{\lambda C\setminus C}d^2x\,\phi(x).
\eeq
We note that the part of the action lying in $\lambda C\setminus C$ gives rise, in the path-integral formalism for instance, to the new state $R_\lambda[C]$ on $\p C$ produced by this single iteration of the Markov chain.

Repeating the process infinitely-many times, the final action is
\beq\label{SClam}
	S = S_\C + g\lt(
	\int_C d^2x\,\phi(x) + \sum_{n=1}^\infty
	\lambda^{n(\Delta_\phi-2)}
	\int_{\lambda^n C\setminus \lambda^{n-1}C}d^2x\,\phi(x)
	\rt).
\eeq
In order to obtain a simpler expression, we may take $\lambda=1+\ep$ near to 1. An analysis of this limit shows that, taking a circular region $C$ of radius $r$ for simplicity, the action becomes
\beq\label{SC}
	S = S_\C + g\int d^2x\,{\rm max}(1,(|x|/r)^{\Delta_\phi-2})\,\phi(x).
\eeq

Thus the resulting state $\lim_{N\to\infty} R^N_\lambda[C]$, both in the cases of $\lambda>1$ \eqref{SClam} and $\lambda\to1$ \eqref{SC}, is described by a correction to the coefficient of the lattice term, which, for $\lambda\to1$, increases with a power $\Delta_\phi-2$ at large distances beyond the region $C$. In particular, the bulk measure is {\em never exactly reached}, contrary to the result \eqref{limR} from the field theory analysis of the Markov chain. That is, the small-scale corrections in \eqref{latsca2} have a non-vanishing effect on the stationary state of the Markov chain, for any value of $\lambda$. However, this effect, being proportional to $g$, is non-universal, and thus should provide only small corrections to bulk behaviours, smaller, for instance, than universal corrections produced by conformal boundary conditions 

Let us analyse the effect of this on a one-point function, subtracting the bulk average. By dimensional analysis, the decay of the effect of the region outside $C$ with the distance $\ell$ to the boundary of $C$ will be proportional to
\beq
	\ell^{n(2-\Delta_\phi)-\Delta_\Or}
\eeq
for the non-universal contribution coming from $n$ factors of the correction term. Note that with a conformal boundary, we would expect instead $\ell^{-\Delta_\Or}$, hence this is a stronger decay as $2-\Delta_\phi<0$.

Let us take the Ising model, and consider the energy field $\Or=\varep$. We must analyse the possible values of $\Delta_\phi$. If we insist on rotation invariance and look at corrections in bulk CFT, the possible fields are the non-derivative, spinless Virasoro descendants of the primary fields, $L_{-2}\b L_{-2}{\bf 1}$, $L_{-2}\b L_{-2}\varep$, and $L_{-2}\b L_{-2}\sigma$. The smallest decay, in this case, would be either for $n=1$ with $L_{-2}\b L_{-2}\varep$ (because this is the only one with nonzero overlap with $\varep$), so $\Delta_\phi = 5$ hence $\ell^{-4}$; or $n=2$ with $L_{-2}\b L_{-2}{\bf 1}$, so $\Delta_\phi=4$ hence $\ell^{-5}$; so the former is the smallest decay. If we include derivative terms, with the justification that for a theory on a finite domain derivative terms contribute to boundary effects, then we may have $\p\b\p\varep$ and $\p\b\p\sigma$. The former gives $\Delta_\phi = 3$ and with $n=1$ this is $\ell^{-2}$.

From the above analysis, we therefore expect that the non-universal effect at distance $\ell$ of the constructed boundary condition be controlled by a power law, which optimally would be $\ell^{-2}$, with a small coefficient.

\end{document}